\documentclass[%
 preprint,
 10pt,
 nofootinbib,
 amsmath,amssymb,
 aps,
 prc,
]{revtex4-1}

\usepackage{graphicx}% Include figure files
\usepackage{dcolumn}% Align table columns on decimal point
\usepackage{bm}% bold math
\usepackage[normalem]{ulem}
\renewcommand\sout{\bgroup \color{red} \ULdepth=-.5ex \ULset}

\begin{document}

\begin{flushright}
KEK-TH-1956
\end{flushright}

\title{
A Study of Degenerate Two-Body and Three-Body Coupled-Channel Systems \\
-Renormalized Effective AGS Equations and Near-Threshold Resonances-
}
\author{
Atsunari Konishi \\
{\it KEK Theory Center, IPNS, \\
High Energy Accelerator Research Organization (KEK), \\
1-1 Oho, Tsukuba, Ibaraki, 205-0801, Japan} \\
\vspace{0.5cm}
Osamu Morimatsu \\
{\it KEK Theory Center, IPNS, \\
High Energy Accelerator Research Organization (KEK), \\
1-1 Oho, Tsukuba, Ibaraki, 205-0801, Japan \\
Department of Physics, Faculty of Science, University of Tokyo, \\
7-3-1 Hongo Bunkyo-ku Tokyo 113-0033, Japan and \\
Department of Particle and Nuclear Studies, \\
Graduate University for Advanced Studies (SOKENDAI), \\
1-1 Oho, Tsukuba, Ibaraki 305-0801, Japan
} \\
\vspace{0.5cm}
Shigehiro Yasui \\
{\it Department of Physics, Tokyo Institute of Technology, \\
2-12-1 Ohokayama, Meguro, Tokyo, 152-8551, Japan \\
%\vspace{0.5cm}
%Preprint Number; KEK TH-No.1956
}
\vspace{0.5cm}
}
\affiliation{}
\date{\today}
\begin{abstract}
Motivated by the existence of candidates for exotic hadrons whose masses are close to both of two-body and three-body hadronic thresholds lying close to each other,
we study degenerate two-body and three-body coupled-channel systems.
We first formulate the scattering problem of non-degenerate two-body and three-body coupled-channels as an effective three-body problem, i.e.\ effective Alt-Grassberger-Sandhas (AGS) equations.
We next investigate the behavior of $S$-matrix poles near the threshold when two-body and three-body thresholds are degenerate.
We solve the eigenvalue equations of the kernel of AGS equations instead of AGS equations themselves to obtain the $S$-matrix pole energy.
We then face a problem of unphysical singularity:
though the physical transition amplitudes have physical singularities only, the kernel of AGS equations have unphysical singularities.
We show, however, that these unphysical singularities can be removed by appropriate reorganization of the scattering equations and mass renormalization.
The behavior of $S$-matrix poles near the degenerate threshold is found to be universal in the sense that the complex pole energy, $E$, is determined by a real parameter, $c$, as $c + E \log{\left( - E \right)} = 0$, or equivalently, ${\rm Im} \, E = \pi {\rm Re} \, E / \log \left({\rm Re} \, E\right)$.
This behavior is different from that of either two-body or three-body system and is characteristic in the degenerate two-body and three-body coupled-channel system.
We expect that this new class of universal behavior might play a key role in understanding exotic hadrons.
\end{abstract}
\maketitle

\section{Introduction}
	The $X\left( 3872 \right)$ was first observed in 2003 \cite{Choi:2003ue} which is considered not to be a simple charmonium \cite{Barnes:2003vb, Choi:2011fc, Hanhart:2011tn} and is therefore a candidate for the exotic hadron. Since its mass is very close to neutral $D {\bar D^{\ast}}$ threshold, it is pointed out that it has significant $D {\bar D^{\ast}}$ molecular component \cite{1976JETPL..23..333V, Tornqvist:1991ks, Swanson:2003tb, Tornqvist:2004qy, Voloshin:2004mh, Braaten:2005ai, Gamermann:2007fi, Liu:2007bf, Liu:2008fh, Thomas:2008ja, Dong:2008gb, Gamermann:2009uq, Close:2009ag, Close:2010wq, Yasui:2013tsa, Yamaguchi:2014hwa}. However, we must be aware that its mass is also very close to $D {\bar D} \pi$ three-body threshold. We therefore have to consider $D {\bar D^{\ast}} \mathchar`- D {\bar D} \pi$ hadronic two-body and three-body coupled-channels analysis whose importance has also been discussed \cite{Suzuki:2005ha, Fleming:2007rp, Braaten:2007ct, Filin:2010se, Baru:2011rs, Kalashnikova:2012qf, Braaten:2015tga}. See \cite{Brambilla:2010cs, Olsen:2015rpa, Chen:2016qju, Hosaka:2016pey} for recent review articles of heavy quarkonium and candidates for the exotic hadron in that energy regions.
	%\item
	%There are both of experimental and theoretical indication that strange dibaryon exists in $K^- pp$ system \cite{Agnello:2005qj, Yamazaki:2010mu, Muramatsu:2012np, Hashimoto:2014cri, Ichikawa:2014rva}. Since the resonance $\Lambda \left( 1405 \right)$ exists in ${\bar K}N$ channel, one need to consider ${\bar K}NN-\Lambda \left( 1405 \right)N-\pi\Sigma N$ coupled-channels analysis might play an important role. See \cite{Feliciello:2015dua, Gal:2016boi} for review articles.
	%\item
	There are both of experimental and theoretical indication that strange dibaryon exists in $K^- pp$ system \cite{Jaffe:1976yi, Feliciello:2015dua, Gal:2016boi}. Since the resonance $\Lambda \left( 1405 \right)$ exists in ${\bar K}N$ channel, ${\bar K}NN\mathchar`-\Lambda \left( 1405 \right)N\mathchar`-\pi\Sigma N$ two-body and three-body coupled-channels effect regarding $\Lambda \left( 1405 \right)$ as a compact $qqq$ baryon 
%, that is, explicit degrees of freedom 
might play an important role in understanding the system.
	%\item
	%Existence of non-strange dibaryons was first discussed in the 60's \cite{Oakes:1963zza, Dyson:1964xwa} just several month after Gell-Mann's first publication on the quark model of hadrons \cite{GellMann:1964nj}.
	Existence of non-strange dibaryons was first discussed in the 60's \cite{Oakes:1963zza, Dyson:1964xwa}, however it is only recent that they are actually observed experimentally \cite{Bashkanov:2008ih, Adlarson:2011bh, Adlarson:2012fe, Adlarson:2012au, Adlarson:2014xmp, Adlarson:2014pxj, Adlarson:2014ozl}. 
%(${\cal D}_{03} \left( 2370 \right)$ according to the grouping in \cite{Dyson:1964xwa}). 
The importance of hadronic two-body and three-body coupled-channels analysis in these channels has also been discussed \cite{Gal:2013dca, Gal:2014zia}. See, for example, an introduction in \cite{Platonova:2014rza} for a review
 of current status of non-strange dibaryon physics. \par
	%\item
	%A recently-found dibaryon resonance ${\cal D}_{03} \left( 2380 \right)$ \cite{Bashkanov:2008ih, Adlarson:2011bh, Adlarson:2012fe, Adlarson:2012au, Adlarson:2014xmp, Adlarson:2014pxj, Adlarson:2014ozl}. This was found in two-pion production process and may not be adequate. Dibaryon including others \cite{Platonova:2014rza}. \lq\lq the role of dibaryon d.o.f in short-range $NN$ correlations''. \par
	%\item
	%Super narrow dibaryon lying below the pion-production threshold \cite{Fil'kov:2013oxa}.
	%\item
	%Possible dibaryon near $N\Delta$ threshold ${\cal D}_{12} \left( 2150 \right)$ in $NN \mathchar`- \pi NN$ processes \cite{Hoshizaki:1978ty, Hoshizaki:1979fu, Hoshizaki:1993ja, Hoshizaki:1993im, Bhandari:1980mu, Arndt:1992kz, Arndt:1993nd, Kravtsov:1984fw}.
	%\item
	Two-body and three-body coupled-channels analysis is therefore required to deepen our understandings of those resonances whose thresholds lie close to each other. In this paper, motivated by such circumstances, we develop two-body and three-body coupled-channels scattering equations and investigate the $S$-matrix pole behavior near the thresholds in case of a degenerate two-body and three-body coupled-channels system. \par
The discussion so far has been focused on phenomenological aspects of hadron physics. However, the degenerate two-body and three-body coupled-channels system is also interesting from a purely theoretical perspective. It is known that the $S$-matrix pole behavior near the threshold in a single-channel two-body and three-body system has the universal property \cite{MR1947260}. Namely, it is determined by one or two parameters depending on how close the poles are located to the threshold. It is also known that universal behavior crucially depends on phase space property near the threshold, that is, whether it is a two-body system, a three-body system or the system has relative angular momentum excitation \cite{Matsuyama:1991bm}. Then, a question arise, \lq\lq How does the $S$-matrix pole behave near the thresholds in case of a degenerate two-body and three-body coupled system?''. We expect a new class of universal behavior emerges in such a case. We might also want to ask the same question in relation to the Efimov effect. If two paris of three particles develop zero-energy bound state, infinite number of bound state appear and is known as the Efimov effect. Degenerate two-body and three-body coupled-channels system corresponds to a three-body system in which {\it one} of three pair develops zero-energy bound state. \par
The behavior we are going to discuss is therefore interesting in its own and also might play a key role in understanding those observed candidates for the exotic hadrons lying in the energy regions where two-body and three-body hadronic thresholds rest close to each other. In this article, we investigate and answer those questions mentioned above. \par
%\vspace{1cm}
In section \ref{setup}, we present basic setups namely, Hamiltonian we consider, effective interactions constructed by the Feshbach projection, the AGS equations which three-body transition amplitudes satisfy, a problem of unphysical singularity and its solution with the mass renormalization plus an appropriate reorganization of the Feynman diagrams. In section \ref{numericalresults}, we calculate the $S$-matrix pole behavior near the thresholds in a degenerate two-body and three-body coupled-channels system using Yamaguchi-type separable interactions. We show that the $S$-matrix pole behavior is characteristic in the system and also universal in a sense that it is determined by the equation $c + E \log{\left( - E \right)} = 0$, or equivalently, ${\rm Im} \, E = \pi {\rm Re} \, E / \log {\rm Re} \, E$, where $E$ is the $S$-matrix pole energy,
while $c$ a real parameter. In section \ref{summaryanddiscussion}, we summarize the results and discuss its physical applications.
%\end{itemize}

\section{Effective AGS equations for two-body and three-body coupled-channels}
\label{setup}

\begin{figure}[htbp]
	\includegraphics[width=8cm]{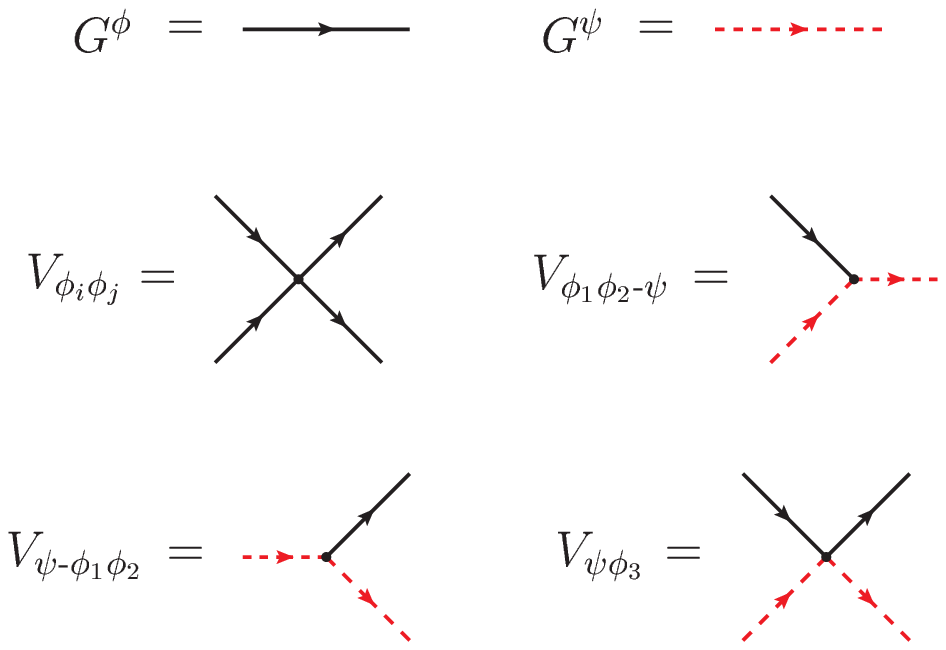}
	\caption{Summary of Feynman rules. We denote propagators of $\phi_i$ as solid lines. Unrenormalized propagator of $\psi$ is represented as red dotted line while the renormalized one as red dashed line (colored online).}
	\label{fig:FeynmanRule}
\end{figure}

We consider a two-body and three-body coupled-channels system. We denote three particles in the three-body channel as $\phi_1\phi_2\phi_3$ and two particles in the two-body channel as $\psi\phi_3$.
Introducing the projection operators onto the three-body channel, $P$,  and the two-body channel, $Q$, respectively,  we write the full Hamiltonian as a matrix \cite{Feshbach:1958nx, Feshbach:1962ut}

\begin{eqnarray}
H = \left(
\begin{array}{cc}
	PHP & PHQ \\
	QHP & QHQ
\end{array} 
\right)
= \left(
\begin{array}{cc}
	H_0^{P} + V_{PP} & V_{PQ} \\
	V_{QP} & H_0^{Q} + V_{QQ}
\end{array} 
\right).
\end{eqnarray}
The kinetic terms in the three-body and two-body channels, $H_0^{\left( 3 \right)}$ and $H_0^{\left( 2 \right)}$, are respectively given by
\begin{eqnarray}
	H_0^{P} &=& \sum_{i=1}^3 \left( m_i + \frac{k_i^2}{2 m_i} \right), \\
	H_0^{Q} &=& M + \frac{K_3^2}{2 M} + m_3 + \frac{k_3^2}{2 m_3}.
\end{eqnarray}
We assume that the diagonal interaction terms in the three-body and two-body channels, $V_{PP}$ and $V_{QQ}$, are given by two-body interactions,
while the off-diagonal interaction terms, $V_{PQ}$ and $V_{QP}$, are due to $\phi_1\phi_2$-$\psi_3$ coupling,
\begin{eqnarray}
\begin{array}{l}
	V_{PP} = V_{\phi_2\phi_3} + V_{\phi_3\phi_1} + V_{\phi_1\phi_2}, \\
	V_{QQ} = V_{\psi\phi_3}, \\
	V_{PQ} = V_{\phi_1\phi_2\mathchar`-\psi}, \\
	V_{QP} = V_{\psi\mathchar`-\phi_1\phi_2}.
\end{array}
\end{eqnarray}
We denote the propagator of $\phi_i$ as $G^{\phi_i}$ defined by
\begin{equation}
	G^{\phi_i} \left( E \right) = \frac{1}{E - m_i - \frac{k_i^2}{2 m_i}},
\end{equation}
and that of $\psi$ as $G^{\psi}$ defined by
\begin{equation}
	G^{\psi} \left( E \right) = \frac{1}{E - M - \frac{K_3^2}{2 M}}.
\end{equation}
The Feynman rules are therefore given as shown in Fig. \ref{fig:FeynmanRule}.

In the following, we adopt  separable interactions both for diagonal and off-diagonal interaction terms, which simplifies the numerical calculation, but the formal argument can be generalized for any interactions. \par

\begin{figure}[htbp]
	\includegraphics[width=11cm]{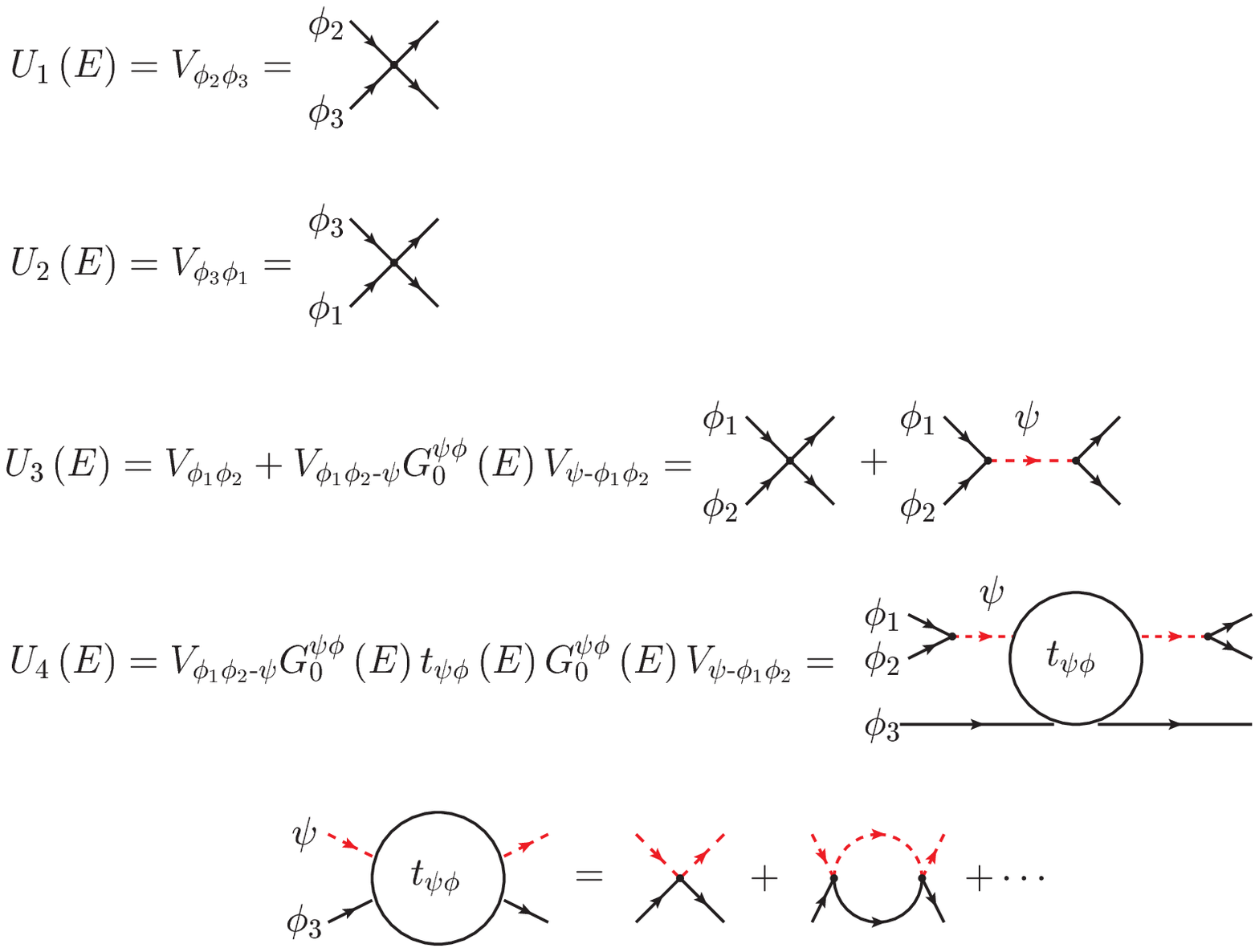}
	\caption{A diagrammatic representation of the effective interactions.}
	\label{fig:EffectiveInteractions}
\end{figure}

We define the effective Hamiltonian in the three-body channel by the Feshbach projection \cite{Feshbach:1958nx, Feshbach:1962ut} as 
\begin{equation}
	H_{eff}^{\left( 3 \right)} = PHP + PHQ \frac{1}{E-QHQ}QHP = H^{P}_0 + U_{PP} \left( E \right),
\end{equation}
where we introduced the effective interaction in the three-body channel, $U_{PP}$, defined by
\begin{equation}
	U_{PP} \left( E \right) = V_{PP} + V_{PQ} \frac{1}{E - H_0^{Q} - V_{QQ}} V_{QP}.
\end{equation}
$U_{PP}$ can be decomposed of the sum of the two-body interactions in each channels, $U_i$ $(i=1,2,3)$, and the three-body interaction, $U_4$:

\begin{equation}
	U_{PP} \left( E \right) = U_1 \left( E \right) + U_2 \left( E \right) + U_3 \left( E \right) + U_4 \left( E \right),
\end{equation}
where each effective interaction is diagrammatically represented as shown in Fig. \ref{fig:EffectiveInteractions}.
\par
The effective interactions are, in a word, sums of scattering processes which make transition {\it once} to the two-body state and make transition back to the three-body state. Scattering processes which make transitions more than once are generated when the effective interactions are iterated. \par
We can see that even in the absence of elementary three-body force, the coupling to the two-body channel generates the effective three-body force. We can also see that the coupling to the two-body channel generates the effective two-body interaction between $\phi_1\phi_2$ in addition to the elementary interaction between them which we denote as $V_3$. \par
\begin{figure}[htbp]
	\includegraphics[width=10cm]{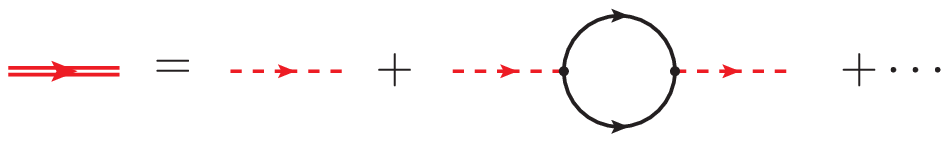}
	\caption{A diagrammatic representation of $G^{\psi} \left( E \right)$.}
	\label{fig:FullPropagator}
\end{figure}

The physical mass of $\psi_3$ is shifted from the bare one by the coupling to $\phi_1\phi_2$.
The dressed Green function of $\psi$, $G^{\psi} \left( E \right)$, is expressed by the free Green function, $G_0^{\psi} \left( E \right)$, and the self energy, $\Sigma \left( E \right)$, as

\begin{equation}
	G^{\psi} \left( E \right) = G_0^{\psi} \left( E \right) + G_0^{\psi} \left( E \right) \Sigma \left( E \right) G_0^{\psi} \left( E \right) + \cdots,
\end{equation}
\begin{equation}
	\frac{1}{E - M - \Sigma \left( E \right)} = \frac{1}{E - M} + \frac{1}{E - M} \Sigma \left( E \right) \frac{1}{E - M} + \cdots,
\end{equation}
which is diagrammatically represented as shown in Fig. \ref{fig:FullPropagator}. The physical mass of $\psi$, $M'$, is determined from the pole energy of the Green function
\begin{equation}
	M' = M + \Sigma \left( E \right) |_{E = M'}.
\end{equation}
In the following, we do not consider the bare interaction which we denote $V_{\phi_1\phi_2}$ for simplicity.

\begin{widetext}
\begin{figure}[htbp]
	\includegraphics[width=15cm]{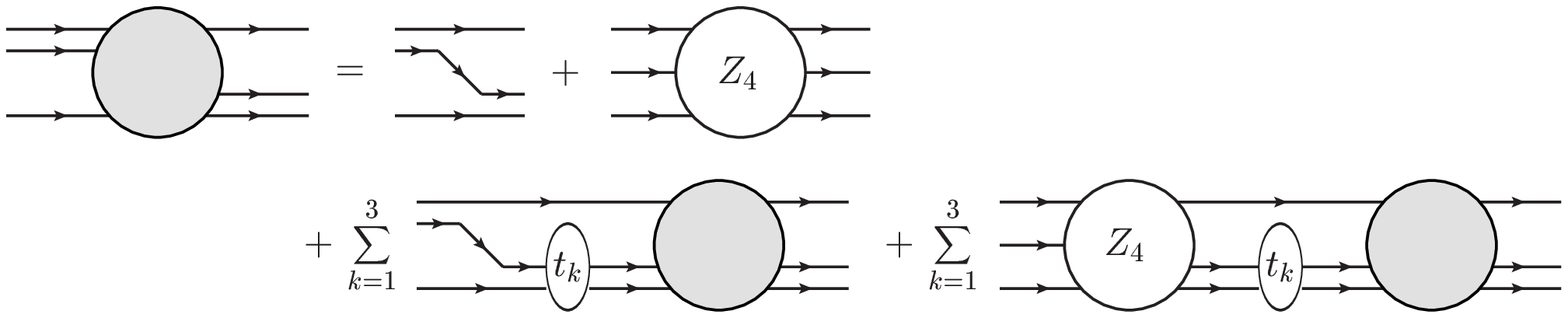}
	\caption{A diagrammatic representation of the AGS equations with three-body force.}
	\label{fig:FaddeevAGS}
\end{figure}
\end{widetext}

Now we want to solve the effective three-body problem defined above.
The three-body problem can be solved by the following AGS equations \cite{Alt:1967fx, Glockle:1983zz}
\begin{equation}
	X\left( E \right) = Z \left( E \right) + Z \left( E \right) T \left( E \right) X \left( E \right),
\end{equation}
where each quantities are $3 \times 3$ matrices whose rows and columns correspond to channel states.
$X \left( E \right)$ is the transition amplitude matrix, $T \left( E \right)$ is a diagonal matrix,
\begin{eqnarray}
T \left( E \right) = \left(
\begin{array}{ccc}
	t_1 \left( E \right) & 0 & 0 \\
	0 & t_2 \left( E \right) & 0 \\
	0 & 0 & t_3 \left( E \right)
\end{array}
\right),
\end{eqnarray}
whose diagonal matrix elements, $t_i (i=1,2,3)$, are the two-body $T$-matrices in each channel,
\begin{equation}
	t_i \left( E \right) = U_i \left( E \right) \frac{1}{1 - G_0^{\phi\phi} \left( E \right) U_i \left( E \right)} \hspace{1cm}( i = 1, 2, 3),
\end{equation}
where $G_0^{\phi\phi} \left( E \right)$ is $\phi\phi$ two-body Green function. $Z \left( E \right)$ is composed of two parts
\begin{equation}
	Z\left( E \right) = Z_0 \left( E \right) + Z_4 \left( E \right).
\end{equation}
$Z_0 \left( E \right)$ has only off-diagonal elements
\begin{eqnarray}
Z_0 \left( E \right) = G_0^{\phi\phi\phi} \left( E \right) {\bar \delta},
\end{eqnarray}
where
\begin{eqnarray}
{\bar \delta} = \left(
\begin{array}{ccc}
	0 & 1 & 1 \\
	1 & 0 & 1 \\
	1 & 1 & 0
\end{array} 
\right),
\end{eqnarray}
and $G_0^{\phi\phi\phi} \left( E \right)$ is $\phi_1\phi_2\phi_3$ three-body Green function. $Z_4 \left( E \right)$ has all $3 \times 3$ components and is a sum of repeated effective three-body force.
\begin{eqnarray}
	Z_4 \left( E \right) & = & \left( G_0^{\phi\phi\phi} \left( E \right) U_4 \left( E \right) G_0^{\phi\phi\phi} \left( E \right) + \cdots \right) {\bf 1} \nonumber \\
	& = & G_0^{\phi\phi\phi} \left( E \right) U_4 \frac{1}{1 - G_0^{\phi\phi\phi} \left( E \right) U_4 \left( E \right)} G_0^{\phi\phi\phi} \left( E \right) {\bf 1},
\end{eqnarray}
where we defined the matrix ${\bf 1}$ as
\begin{eqnarray}
{\bf 1} = \left(
\begin{array}{ccc}
	1 & 1 & 1 \\
	1 & 1 & 1 \\
	1 & 1 & 1
\end{array} 
\right).
\end{eqnarray}
The off-diagonal structure of $Z_0 \left( E \right)$ combined with diagonal nature of $T \left( E \right)$ prevents overcounting the same two-body $T$-matrices in a row. A diagrammatic representation of the scattering equations is given in Fig. \ref{fig:FaddeevAGS}. The diagrammatic representation clearly shows what is done in the scattering equations. We first sum a three-body interaction in addition to three two-body interactions to give the three two-body $T$-matrices $t_i\ \left( i = 1, 2, 3 \right)$ and $Z_4 \left( E \right)$. We then sum
%the four $T$-matrices
 them up mixing with each other while taking care of overcounting the same two-body $T$-matrices in a row.
The solution of the AGS equations is formally given as
\begin{equation}
	X \left( E \right) = \frac{1}{1 - Z \left( E \right) T \left( E \right)} Z \left( E \right).
\end{equation}
Consider the eigenvalue equation of the kernel of the AGS equations
\begin{equation}
	Z \left( E \right) T \left( E \right) | n \rangle = \eta_n \left( E \right) | n \rangle,
	\label{eq:EigenvalueEquation}
\end{equation}
with the eigenvector and eigenvalue, the formal solution is written as
\begin{equation}
	X \left( E \right) = \sum_n \frac{| n \rangle \langle n |}{1 - \eta_n \left( E \right)} Z \left( E \right).
	\label{eq:XEigenvalueRepresentation}
\end{equation}
If it has an eigenvalue, 1, at the energy $E = E_p$, i.e. $\eta_n \left( E_p \right) = 1$, then $X\left( E \right)$ has a pole at the energy $E = E_p$ as can be seen from Eq.(\ref{eq:XEigenvalueRepresentation}). Therefore, we solve the eigenvalue equation of the kernel of the effective AGS equations instead of solving the equations themselves. \par
However, if we na{\" i}vely try to solve the eigenvalue equation Eq.(\ref{eq:EigenvalueEquation}), we face the problem of an {\it unphysical singularity} since a part of the driving term $Z_4 \left( E \right)$ has unphysical singularities though the transition amplitudes $X \left( E \right)$ has only physical singularities, which is discussed in detail in the following.

We first group the transition amplitudes into the third channel $3$, $t_3 \left( E \right)$, and the effective three-body interaction. To that end, we decompose the two-body $T$-matrix as
\begin{equation}
	T \left( E \right) = {\bar T_3} \left( E \right) + T_3 \left( E \right),
\end{equation}
where written in an explicit matrix form,
\begin{eqnarray}
{\bar T_3} \left( E \right) = \left(
\begin{array}{ccc}
	t_1 \left( E \right) & 0 & 0 \\
	0 & t_2 \left( E \right) & 0 \\
	0 & 0 & 0 \\
\end{array} 
\right), \hspace{1cm}
T_3 \left( E \right) = \left(
\begin{array}{ccc}
	0 & 0 & 0 \\
	0 & 0 & 0 \\
	0 & 0 & t_3 \left( E \right) \\
\end{array}
\right),
\end{eqnarray}
and rewrite the $X \left( E \right)$ as follows
\begin{equation}
	X \left( E \right) = X_3 \left( E \right) + X_3 \left( E \right) {\bar T_3} \left( E \right) X_3 \left( E \right) + \cdots = \frac{1}{1 - X_3 \left( E \right) {\bar T_3} \left( E \right)} X_3 \left( E \right),
\end{equation}
where
\begin{equation}
	X_3 \left( E \right) = Z \left( E \right) + Z \left( E \right) T_3 \left( E \right) Z \left( E \right) + \cdots.
\end{equation}
Substituting $Z \left( E \right) = Z_0 \left( E \right) + Z_4 \left( E \right)$ and noting that $T_3 \left( E \right) Z_0 \left( E \right) T_3 \left( E \right) = 0$, we can simplify $X_3 \left( E \right)$ as
\begin{equation}
	X_3 \left( E \right) = Z_0 \left( E \right) + Z_0 \left( E \right) T_3 \left( E \right) Z_0 \left( E \right) + \left( 1 + Z_0 \left( E \right) T_3 \left( E \right) \right) W_3 \left( E \right) \left( T_3 \left( E \right) Z_0 \left( E \right) + 1 \right),
\end{equation}
where we define $W_3 \left( E \right)$ by
\begin{equation}
	W_3 \left( E \right) = Z_4 \left( E \right) + Z_4 \left( E \right) T_3 \left( E \right) Z_4 \left( E \right) + \cdots = \frac{1}{1 - Z_4 \left( E \right) T_3 \left( E \right)} Z_3 \left( E \right).
\end{equation}
The matrices, $W_3 \left( E \right)$ and $Z_4 \left( E \right)$ are channel-independent, i.e.\ $W_3 \left( E \right) = w_3 \left( E \right) {\bf 1}$ and $Z_4 \left( E \right) = z_4 \left( E \right) {\bf 1}$. $w_3 \left( E \right)$ is the part in which $t_3 \left( E \right)$ and the effective three-body interaction are collected,
\begin{equation}
	w_3 \left( E \right) = z_4 \left( E \right) + z_4 \left( E \right) t_3 \left( E \right) z_4 \left( E \right) + \cdots = \frac{1}{1 - z_4 \left( E \right) t_3 \left( E \right)} z_4 \left( E \right).
\end{equation}
Diagrammatically, $z_4 \left( E \right)$ is given by
\begin{equation}
	\raisebox{-0.4cm}{\includegraphics[width=13cm]{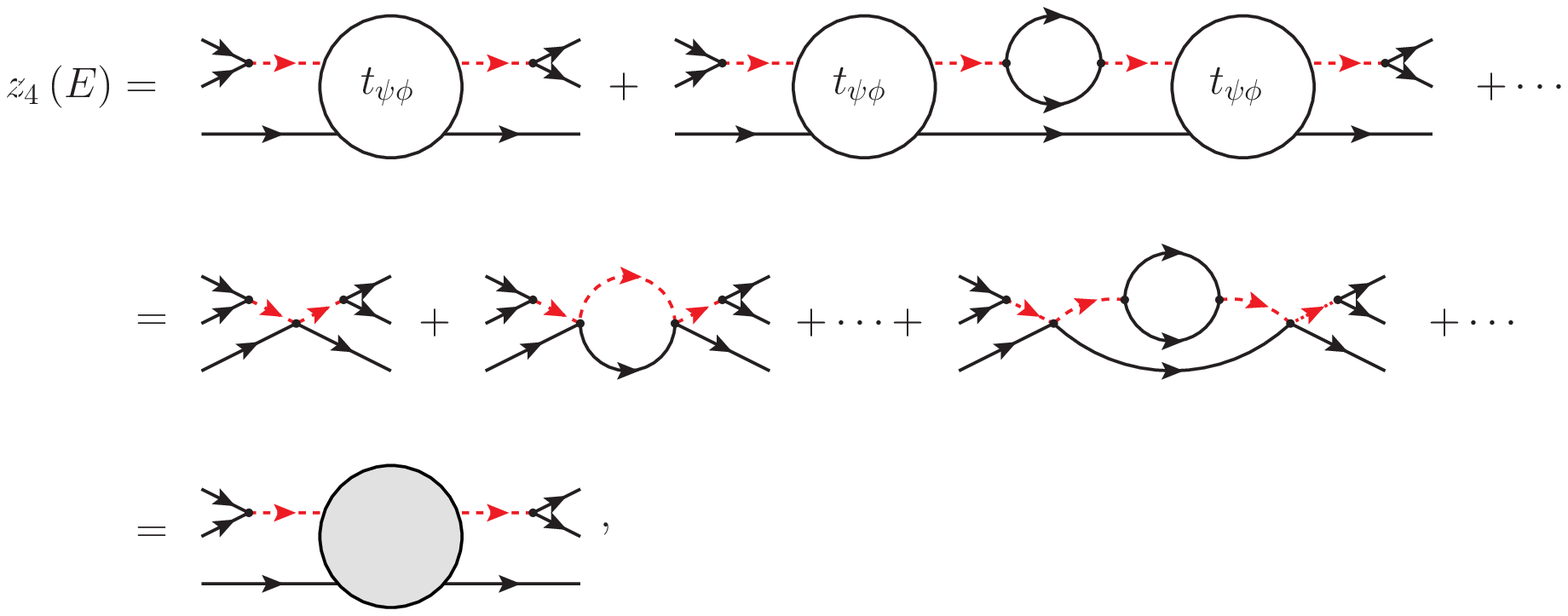}},
\end{equation}
and $z_4 \left( E \right) t_3 \left( E \right) z_4 \left( E \right)$ by
\begin{equation}
	\raisebox{-0.8cm}{\includegraphics[width=15cm]{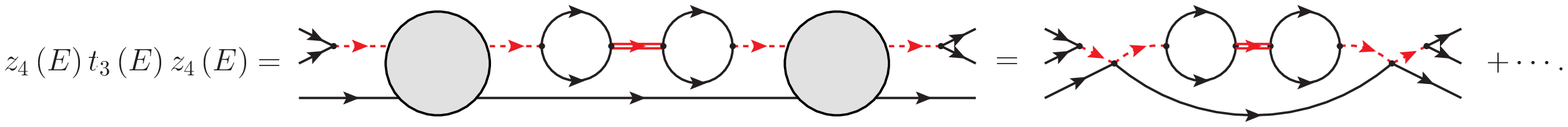}}
\end{equation}
$w_3 \left( E \right)$ is therefore given by
\begin{equation}
	\raisebox{-0.6cm}{\includegraphics[width=11cm]{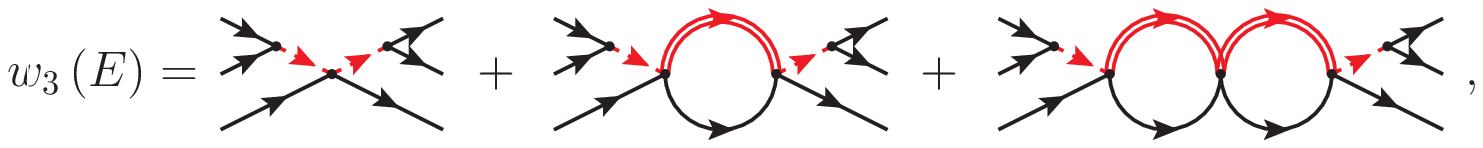}}
\end{equation}
where in the intermediate state the use has been made of the relation for the Green function for $\psi$,
\begin{eqnarray}
	G^{\psi} \left( E \right) &=& G^{\psi}_0 \left( E \right) + G^{\psi}_0 \left( E \right) \Sigma \left( E \right) G^{\psi}_0 \left( E \right) + G^{\psi}_0 \left( E \right) \Sigma \left( E \right) G^{\psi} \left( E \right) \Sigma \left( E \right) G^{\psi}_0 \left( E \right),
	\label{eq:FullPropagator}
\end{eqnarray}
or diagrammatically,
\begin{equation}
	\raisebox{-0.8cm}{\includegraphics[width=10cm]{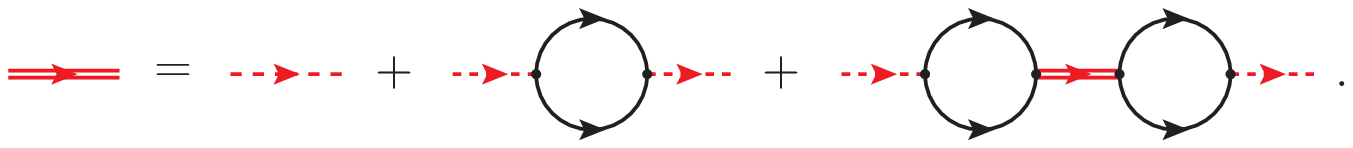}}
\end{equation}
The first two terms in Eq.(\ref{eq:FullPropagator}), $G^{\psi}_0 \left( E \right)$ and $G^{\psi}_0 \left( E \right) \Sigma \left( E \right) G^{\psi}_0 \left( E \right)$, are included in $z_4 \left( E \right)$,
while the last term $G^{\psi}_0 \left( E \right) \Sigma \left( E \right) G^{\psi} \left( E \right) \Sigma \left( E \right) G^{\psi}_0 \left( E \right)$ in $z_4 \left( E \right) t_3 \left( E \right) z_4 \left( E \right)$,
and the sum of these terms gives the full Green function, $G^{\psi} \left( E \right)$, in $w_3 \left( E \right)$.

Therefore, it is clear that $z_4 \left( E \right)$ as well as  $Z_4 \left( E \right)$ has an unphysical branch point at $M+m_3$, while $w_3 \left( E \right)$ as well as $W_3 \left( E \right)$ has only a physical one at $M'+m_3$.
If we solve for the scattering amplitudes $X \left( E \right)$, this unphysical singularity of $Z \left( E \right)$ does not matter.
However, since we solve the eigenvalue equation of $Z \left( E \right) T \left( E \right)$, which has an unphysical singularity,
\footnote
{This unphysical singularity problem manifests itself in the following simple example.
%The unphysical singularity problem manifests itself in the following simple system.
 We consider a particle in the rest frame and regard a part of the physical mass as an interaction as
\[
	H = m = m_B + \delta m.
\]
The Dyson-Schwinger equation that the full Green function satisfies is
\[
	G \left( E \right) = G_B \left( E \right) + G_B \left( E \right) \delta m G \left( E \right),
\]
where the full and bare Green functions are defined by
\[
	G \left( E \right) = ( E - m )^{-1}, \qquad
	G_B \left( E \right) = ( E - m_B )^{-1}.
\]
Eigenvalue equation of the Dyson-Schwinger equation is
\[
	G_B \left( E \right) \delta m | \phi \rangle = \eta \left( E \right) | \phi \rangle,
\]
whose eigenvalue is obviously
\[
	\eta \left( E \right) = \frac{\delta m}{E - m_B},
\]
and it has an {\it unphysical singularity}.
}
the existence of the unphysical singularity causes difficulty in searching for the S-matrix pole in the complex energy plane.
In order to avoid unphysical singularities we reorganize the above effective AGS equations in terms of the renormalized free Green function.

In order to avoid unphysical singularities we reorganize the above effective AGS equation in terms of the renormalized free Green's function.
We define the renormalized free Hamiltonian as the kinetic term with the physical mass,
\begin{equation}
	{H'_0}^{ Q} = M' + \frac{K^2}{2 M'} + m_3 + \frac{k_3^2}{2 m_3},
\end{equation}
and subtract the difference of the bare and renormalized free Hamiltonians, $\Delta$, from the interaction, $V_{\psi\phi_3}$:
\begin{eqnarray}
	{H'_0}^{Q} &=& H_0^{Q} - \Delta, \\
	V'_{\psi\phi_3} &=& V_{\psi\phi_3} + \Delta.
\end{eqnarray}
Then, we show that in terms of the renormalized free Green function 
\begin{equation}
	{G'_0}^{\psi} \left( E \right) = \frac{1}{ E - M' - \frac{K_3^2}{2 M'}}.
\end{equation}
We can reorganize the effective AGS equations as
\begin{eqnarray}
	X' \left( E \right) &=& Z' \left( E \right) + Z' \left( E \right) T \left( E \right) X' \left( E \right), \\
	Z' \left( E \right) &=& Z_0 \left( E \right) + Z'_4 \left( E \right),
\end{eqnarray}
where the modified kernel, $Z' \left( E \right)$, has only physical singularities.
\par
Let us first note that the relation for the Green function, Eq.(\ref{eq:FullPropagator}), is modified in terms of the renormalized free Green's function, ${G'_0}^\psi \left( E \right)$ as
\begin{eqnarray}
	G^\psi \left( E \right) &=& {G'_0}^\psi \left( E \right) + {G'_0}^\psi \left( E \right) (\Sigma \left( E \right) + \Delta) {G'_0}^\psi \left( E \right) + {G'_0}^\psi \left( E \right) (\Sigma \left( E \right) + \Delta) G^\psi \left( E \right) (\Sigma \left( E \right) - \Delta) {G'_0}^\psi \left( E \right) \nonumber \\
	&=& {G'_0}^\psi \left( E \right) + {G'_0}^\psi \left( E \right) \Sigma \left( E \right) {G'_0}^\psi \left( E \right) + {G'_0}^\psi \left( E \right) \Delta {G'_0}^\psi \left( E \right) + {G'_0}^\psi \left( E \right) \Delta G^\psi \left( E \right) \Delta {G'_0}^\psi \left( E \right) \nonumber \\
	& + & {G'_0}^\psi \left( E \right) \Delta G^\psi \left( E \right) \Sigma {G'_0}^\psi \left( E \right) + {G'_0}^\psi \left( E \right) \Sigma G^\psi \left( E \right) \Delta {G'_0}^\psi \left( E \right) + {G'_0}^\psi \left( E \right) \Sigma \left( E \right) G^\psi \left( E \right) \Sigma \left( E \right) {G'_0}^\psi \left( E \right),
\end{eqnarray}
or diagrammatically,
\begin{equation}
	\raisebox{0cm}{\includegraphics[width=17cm]{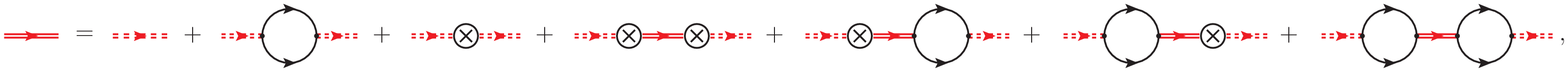}}
\end{equation}
where we represent the renormalized free Green function as double-dashed line as% shown below.
\begin{equation}
	\raisebox{0cm}{\includegraphics[width=3.5cm]{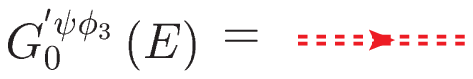}}.
\end{equation}
%\begin{figure}[htbp]
%	\includegraphics[width=3.5cm]{FeynmanRules_2.eps}.
%	\caption{A diagrammatic representations of the renormalized free Green function.}
%\end{figure}
In terms of the renormalized free Green function, we define $Z_4' \left( E \right)$ by
\begin{equation}
	\raisebox{-0.4cm}{\includegraphics[width=17cm]{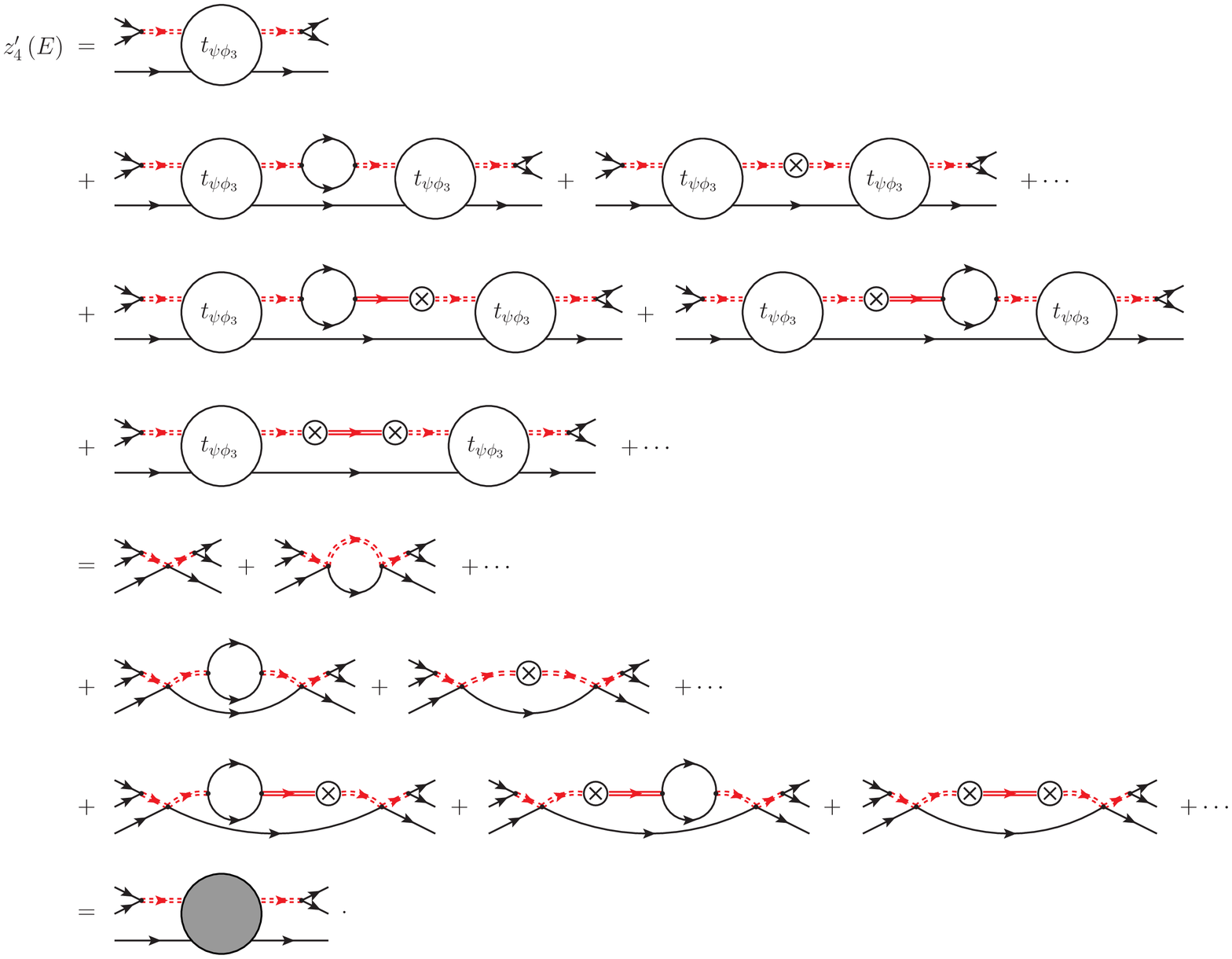}}
\end{equation}
and then $z'_4 \left( E \right) t_3 \left( E \right) z'_4 \left( E \right)$ becomes
\begin{equation}
	\raisebox{-0.4cm}{\includegraphics[width=15cm]{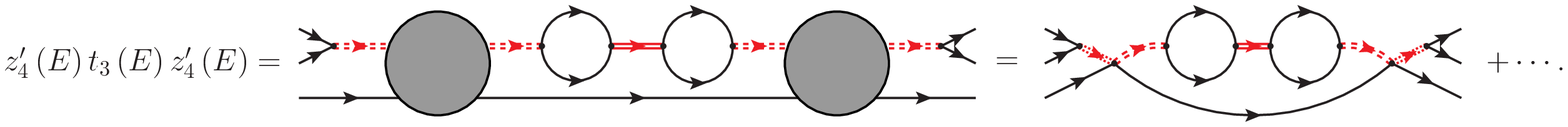}}
\end{equation}
Recalling that the corresponding counterterms are included in $z_4' \left( E \right)$, $w_3' \left( E \right)$ is therefore given by
\begin{equation}
	\raisebox{-0.4cm}{\includegraphics[width=10cm]{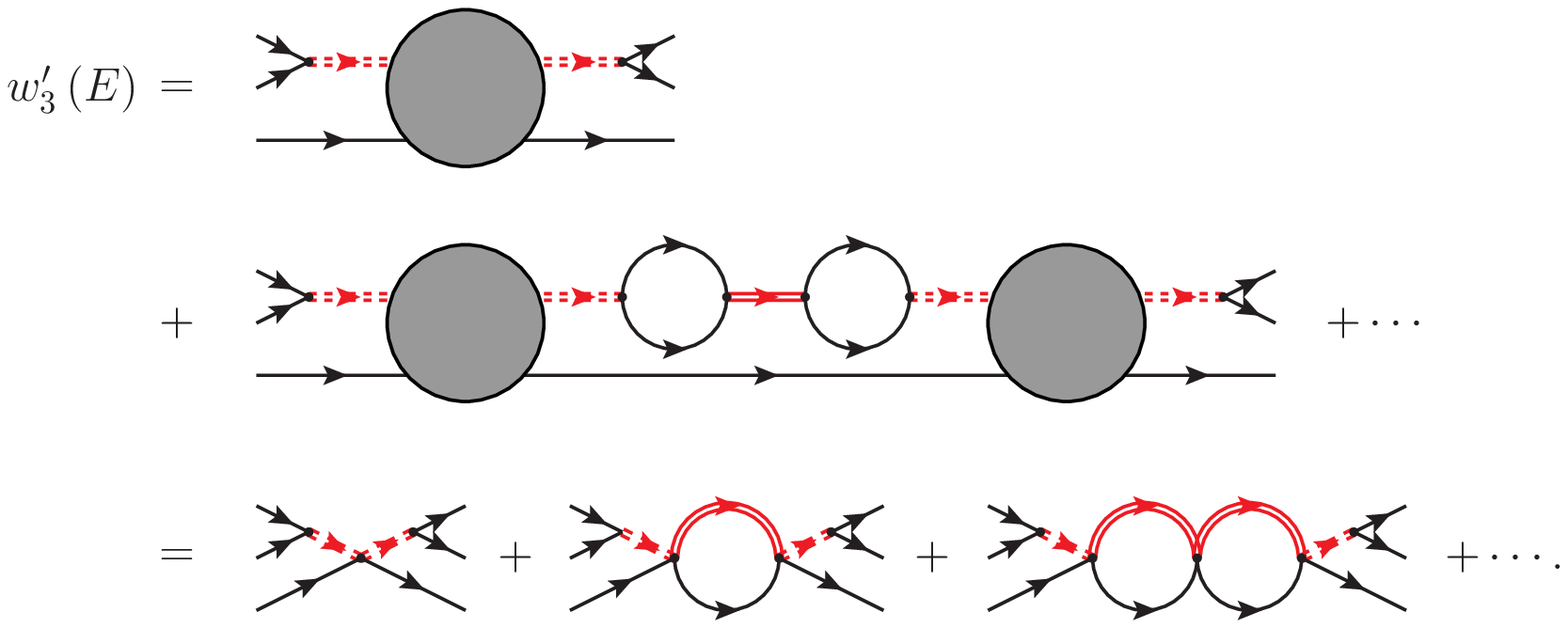}}
\end{equation}
Clearly, $w'_3 \left( E \right)$ is the same as $w_3 \left( E \right)$ (except for the free Green functions at both ends).
Yet, given as a finite sum of the renormalized free Green's function, $z'_4 \left( E \right)$ has a physical branch point at $M' + m_3$ not an unphysical one at $M + m_3$ in contrast to $z_4 \left( E \right)$.
This is because by expressing in terms of the renormalized free Green's function, $z'_4 \left( E \right)$ includes not only $z_4 \left( E \right)$ but also a part of the higher order terms, $z_4 \left( E \right) t_3 \left( E \right) z_4 \left( E \right)$, $z_4 \left( E \right) t_3 \left( E \right) z_4 \left( E \right) t_3 \left( E \right) z_4 \left( E \right)$, and so on, which changed the bare singularity to the physical one. \par
Channel-dependent parts of $Z_4' \left( E \right)$ are given as %shown below
(see appendix \ref{sec:appendix1} for details)
\begin{equation}
	\raisebox{0cm}{\includegraphics[width=12cm]{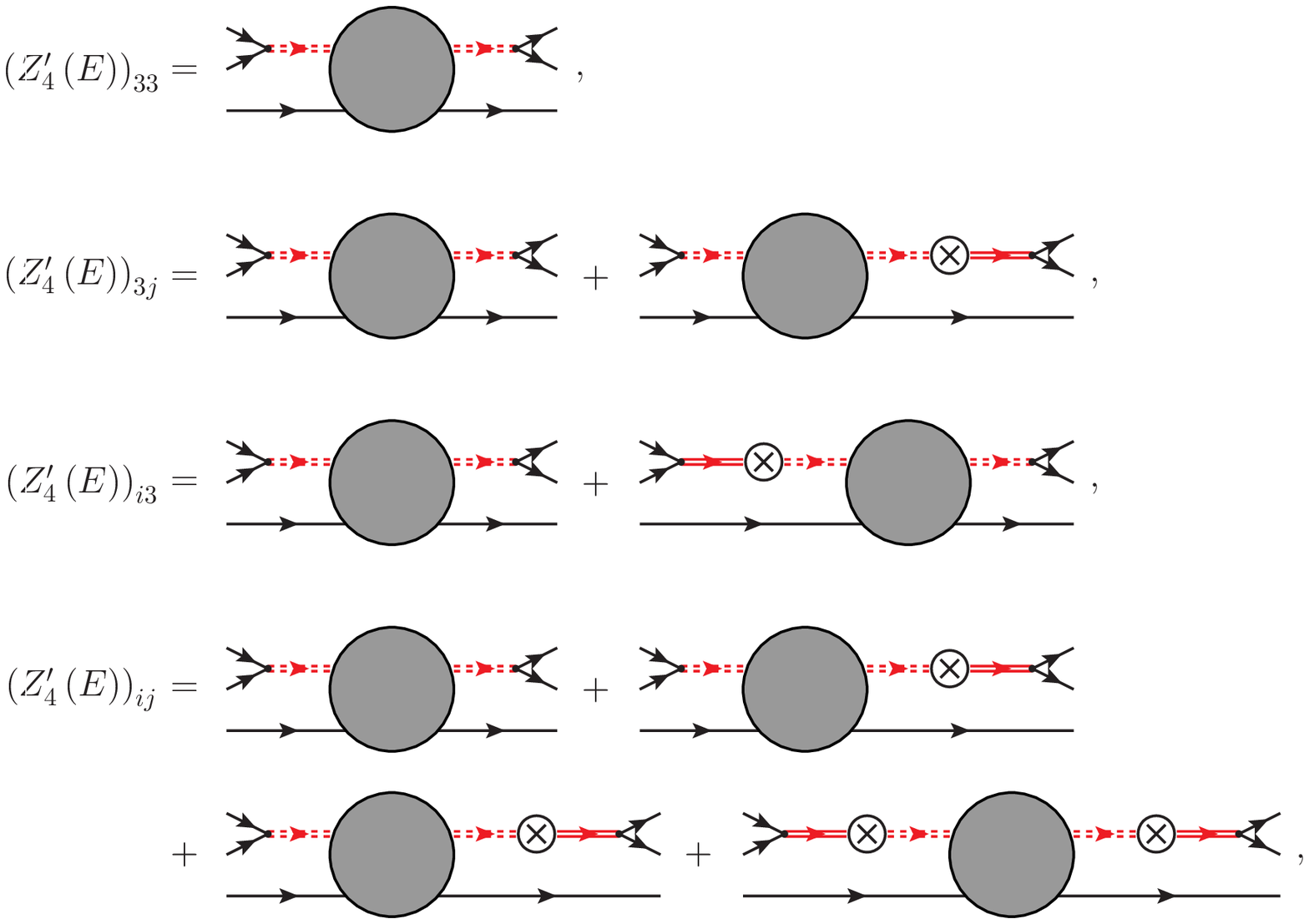}}
\end{equation}
where the blob is defined in $z_4' \left( E \right)$.
%\begin{figure}[htbp]
%	\includegraphics[width=12cm]{Z4_4.eps}
%	\caption{Diagrammatic representation of matrix elements of the modified kernel. The blob is
% given in Eq.(ref equation)
%defined in $z_4' \left( E \right)$.}
%\end{figure}

\section{The $S$-matrix pole behavior near the threshold; numerical results}\label{numericalresults}
In this section, we numerically solve the eigenvalue equation of the kernel of the effective AGS equations and analyze the $S$-matrix pole behavior near the thresholds in case of a degenerate two-body and three-body coupled-channels system.

We adopt separable interactions both for diagonal and off-diagonal parts whose matrix elements are
\begin{eqnarray}
	\langle k_i k_j | V_{\phi_i\phi_j} | k'_i k'_j \rangle &=& g \left( p_k \right) \lambda_{\phi_i\phi_j} g \left( p_k' \right), \\
	\langle K k_3 | V_{\psi\phi_3} | K' k'_3 \rangle &=& g \left( q_3 \right) \lambda_{\psi\phi} g \left( q_3' \right), \\
	\langle K | V_{\psi\mathchar`-\phi_1\phi_2} | k_1 k_2 \rangle &=& \frac{1}{K^2} \delta \left( K - k_1 - k_2 \right) \lambda_{\psi\mathchar`-\phi\phi} g \left( p_3 \right).
\end{eqnarray}
$p_k$ is the relative momentum of $\phi_i$ and $\phi_j$ , $p_k=(k_i-k_j)/2$ while $q_3$ is that of $\psi$ and $\phi_3$, $q_k=(K-k_3)/2$.
$g \left( p \right)$ is Yamaguchi-type form factor
$
	g \left( p \right) = \Lambda^2/({p^2 + \Lambda^2}),
$ where $\Lambda$ is called a cut-off parameter.
$\lambda_{\phi\phi}$ and $\lambda_{\psi\phi}$ are the coupling constants of the two-body interactions in the two-body and three-body channels, respectively.
In the two-body channel we take the coupling constants of the two-body interactions of $\phi_2\phi_3$ and $\phi_3\phi_1$ to be the same, $\lambda_{\phi\phi}$, but that of $\phi_1\phi_2$ to be zero.
$\lambda_{\psi\mathchar`-\phi\phi}$ is $\psi\mathchar`-\phi_1\phi_2$ coupling constant.

%\subsection{Analytic Continuation to the Unphysical Complex Energy Sheet}
Bound state poles lie on the real physical energy axis below the threshold, which can be directly studied by the eigenvalue equation of the kernel of the modified effective AGS equations
%{\color{red}
\begin{equation}
	Z' \left( E \right) T \left( E \right) | n \rangle = \eta_n \left( E \right) | n \rangle.
	\label{eq:ModifiedEigenvalueEquation}
\end{equation}
%}
On the other hand, resonance poles are located in the fourth quadrant of the unphysical energy sheet, which can be exposed only after analytic continuation of the eigenvalue equation Eq.(\ref{eq:ModifiedEigenvalueEquation}) is performed onto the unphysical complex energy sheet.
For this purpose we adopt the method of contour rotation.
See, for example, \cite{Pearce:1984ca, Afnan:1991kb}. \par
The contour rotation is done by rotating the integration contour as follows
\begin{widetext}
\begin{eqnarray}
	& & \sum_{j = 1}^3 \int^{\infty}_0 q_j^2 dq_j Z' \left( E q_i q_j \right) \tau_j \left( E - m_j - \frac{q_j^2}{2 M_j} \right) \phi \left( q_j \right) = \eta \left( E \right) \phi \left( q_i \right), \nonumber \\
	& \Rightarrow & \sum_{j = 1}^3 \int^{\infty}_0 e^{- 3 i \theta} q_j^2 dq_j Z' \left( E e^{- i \theta} q_i e^{- i \theta} q_j \right) \tau_j \left( E - m_j - \frac{e^{- 2 i \theta} q_j^2}{2 M_j} \right) \phi \left( e^{-i \theta} q_j \right) = \eta \left( E \right) \phi \left( e^{- i \theta} q_i \right).
\end{eqnarray}
\end{widetext}
Since we adopt Yamaguchi-type form factor, the rotation angle $\theta$ in the momentum plane is restricted to $\theta \leq \frac{\pi}{2}$ and therefore an angle of rotation of the branch cut in the complex energy sheet is restricted to be less than $\pi$. This causes no problem since we are interested in resonances lying on the fourth quadrant of the unphysical complex energy sheet.

After performing analytical continuation, we two-dimensionally discretize the fourth quadrant of the complex energy sheet
%{\color{red}%
with the interval
%}
$0.00025$ and calculate the Fredholm determinant, $\prod_n \left( 1 - \eta_n \left( E \right) \right)$, of the kernel, $Z \left( E \right) T \left( E \right)$, at each grid.
We then identify the complex energy, at which the Fredholm determinant takes a minimum, as an  approximate pole energy with regarding discretization intervals as errors.
\par

%{\color{red}
 Here we summarize how we take the parameters of the model.
%}
We assume the masses of $\phi_i$ to be the same and take them as the unit of the energy, $m_1 = m_2 = m_3 = 1$.
The physical mass of $\psi$ is therefore two, $M' = 2$, in our unit.
It is useful to define dimensionless coupling constants $f_{\phi\phi}$ and $f_{\psi\phi}$ as
$f_{\phi\phi} = \frac{\pi}{2}\mu_{\phi\phi}\lambda_{\phi\phi}\Lambda$ and $f_{\psi\phi} = \frac{\pi}{2}\mu_{\psi\phi}\lambda_{\psi\phi}\Lambda$ where $\mu_{\phi\phi}$ and $\mu_{\psi\phi}$ are reduced masses of $\phi\phi$ and $\psi\phi$, respectively.
When $f_{\phi\phi} \le -1$ ($f_{\psi\phi} \le -1$) the two-body system, $\phi_2\phi_3$ or $\phi_3\phi_1$ ($\psi\phi_3$) has a bound state. 
Matrix elements of the effective interactions, two-body and three-body $T$-matrices, driving terms are presented in the appendix \ref{sec:appendix2}.
We fix the off-diagonal coupling constant, $\lambda_{\psi\mathchar`-\phi\phi}$, to be $\sqrt{4 \pi}\times 0.1$ in our unit for simplicity,
while we change the diagonal coupling constants, i.e.\ we take six values, $-0.15$, $-0.2$, $-0.25$, $-0.3$, $-0.35$, $-0.4$, for $f_{\phi\phi}$, and ten to fifteen values between $-1.11$ and $-1.41$ for $f_{\psi\phi}$.

\begin{turnpage}
\begin{table}[htbp]
		\begin{tabular}{|c||c|c|c|c|c|c|}
			\hline
			$ f_{3'3}$ / $f_{23} $ & $-0.15$ & $-0.20$ & $-0.25$ & $-0.3$ & $-0.35$ & $-0.4$ \\ \hline
			$-1.11$&                               &                               &                               &                               &$( 0.00035, -0.00990 )$&$( 0.00035, -0.00740 )$\\ \hline
			$-1.13$&                               &                               &                               &$( 0.00110, -0.01115 )$&$( 0.00085, -0.00890 )$&$( 0.00060, -0.00690 )$\\ \hline
			$-1.15$&                               &                               &                               &$( 0.00135, -0.00990 )$&$( 0.00110, -0.00790 )$&$( 0.00085, -0.00615 )$\\ \hline
			$-1.17$&                               &                               &$( 0.00235, -0.01100 )$&$( 0.00160, -0.00890 )$&$( 0.00135, -0.00715 )$&$( 0.00110, -0.00540 )$\\ \hline
			$-1.19$&                               &$( 0.00310, -0.01150 )$&$( 0.00235, -0.00965 )$&$( 0.00185, -0.00790 )$&$( 0.00135, -0.00715 )$&$( 0.00110, -0.00465 )$\\ \hline
			$-1.21$&$( 0.00385, -0.01140 )$&$( 0.00310, -0.00990 )$&$( 0.00260, -0.00840 )$&$( 0.00210, -0.00690 )$&$( 0.00160, -0.00540 )$&$( 0.00110, -0.00390 )$\\ \hline
			$-1.23$&$( 0.00385, -0.00990 )$&$( 0.00310, -0.00865 )$&$( 0.00260, -0.00715 )$&$( 0.00210, -0.00590 )$&$( 0.00160, -0.00440 )$&$( 0.00085, -0.00315 )$\\ \hline
			$-1.25$&$( 0.00385, -0.00865 )$&$( 0.00310, -0.00740 )$&$( 0.00260, -0.00615 )$&$( 0.00185, -0.00490 )$&$( 0.00135, -0.00365 )$&$( 0.00085, -0.00190 )$\\ \hline
			$-1.27$&$( 0.00360, -0.00740 )$&$( 0.00285, -0.00615 )$&$( 0.00235, -0.00515 )$&$( 0.00185, -0.00390 )$&$( 0.00110, -0.00265 )$&$( 0.00110, -0.00140 )$\\ \hline
			$-1.29$&$( 0.00335, -0.00615 )$&$( 0.00260, -0.00515 )$&$( 0.00210, -0.00415 )$&$( 0.00160, -0.00315 )$&$( 0.00110, -0.00165 )$&$( 0.00110, -0.00090 )$\\ \hline
			$-1.31$&$( 0.00285, -0.00490 )$&$( 0.00235, -0.00415 )$&$( 0.00185, -0.00315 )$&$( 0.00135, -0.00215 )$&$( 0.00110, -0.00115 )$&$( 0.00085, -0.00065 )$\\ \hline
			$-1.33$&$( 0.00260, -0.00390 )$&$( 0.00210, -0.00315 )$&$( 0.00160, -0.00215 )$&$( 0.00135, -0.00115 )$&$( 0.00110, -0.00065 )$&$( 0.00085, -0.00040 )$\\ \hline
			$-1.35$&$( 0.00210, -0.00290 )$&$( 0.00160, -0.00215 )$&$( 0.00135, -0.00140 )$&$( 0.00110, -0.00065 )$&$( 0.00085, -0.00040 )$&$( 0.00060, -0.00040 )$\\ \hline
			$-1.37$&$( 0.00160, -0.00190 )$&$( 0.00135, -0.00115 )$&$( 0.00110, -0.00065 )$&$( 0.00085, -0.00040 )$&$( 0.00060, -0.00040 )$&$( 0.00010,  0.00010 )$\\ \hline
			$-1.39$&$( 0.00135, -0.00150 )$&$( 0.00110, -0.00065 )$&$( 0.00085, -0.00040 )$&$( 0.00060, -0.00040 )$&$( 0.00035, -0.00015 )$&                               \\ \hline
			$-1.41$&$( 0.00010,   0.00010 )$&                               &                               &                               &                               &                              \\ \hline
		\end{tabular}
	\caption{Approximate pole positions on the unphysical complex energy sheet on which resonances lie for various coupling constants. Each row corresponds to each $f_{\psi\phi_3}$ while each column corresponds to each $f_{\phi\phi}$.}
	\label{table:poletrajectories}
\end{table}
\end{turnpage}

%\pagebreak
\begin{figure}[htbp]
	\begin{tabular}{cc}
		\begin{minipage}[htbp]{0.5\hsize}
			\centering
			\includegraphics[width=8cm, angle = 0]{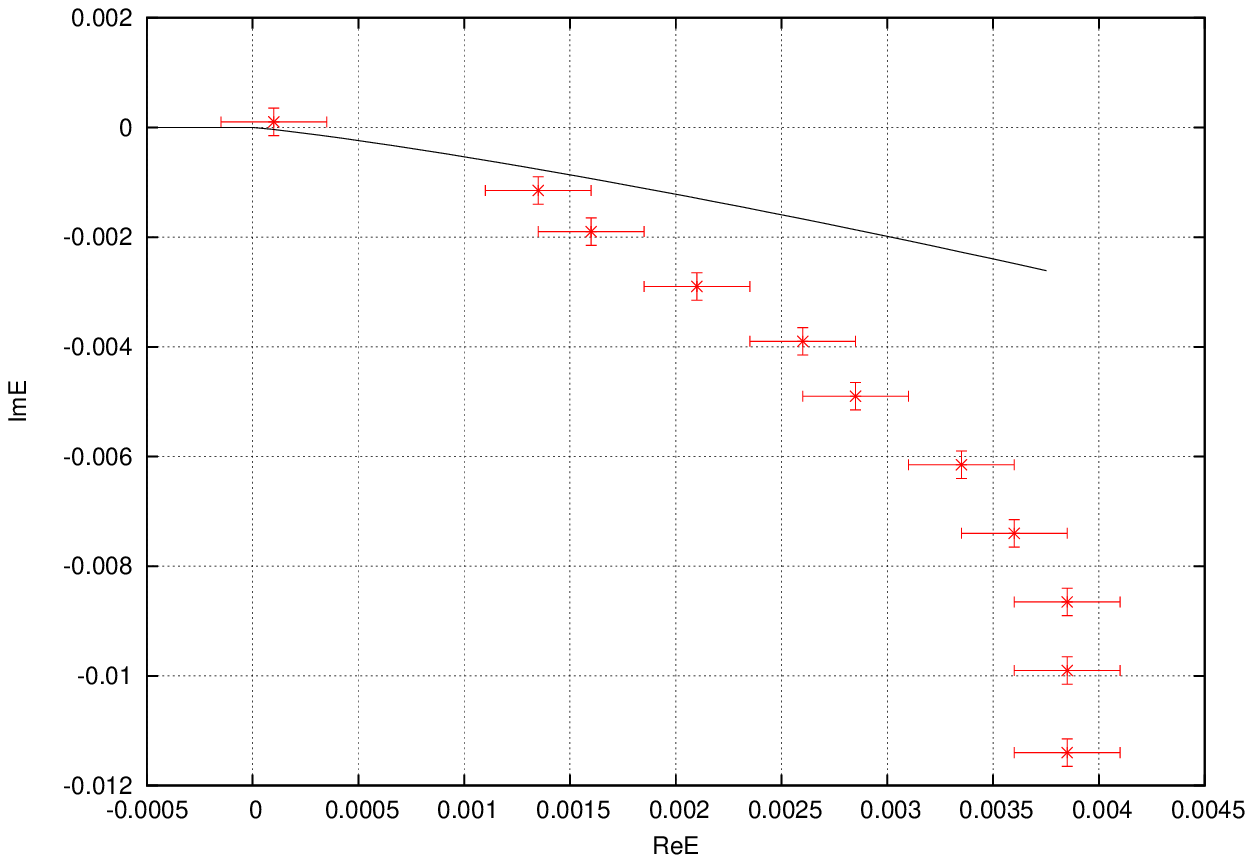}
		\end{minipage} &
		\begin{minipage}[htbp]{0.5\hsize}
			\centering
			\includegraphics[width=8cm, angle = 0]{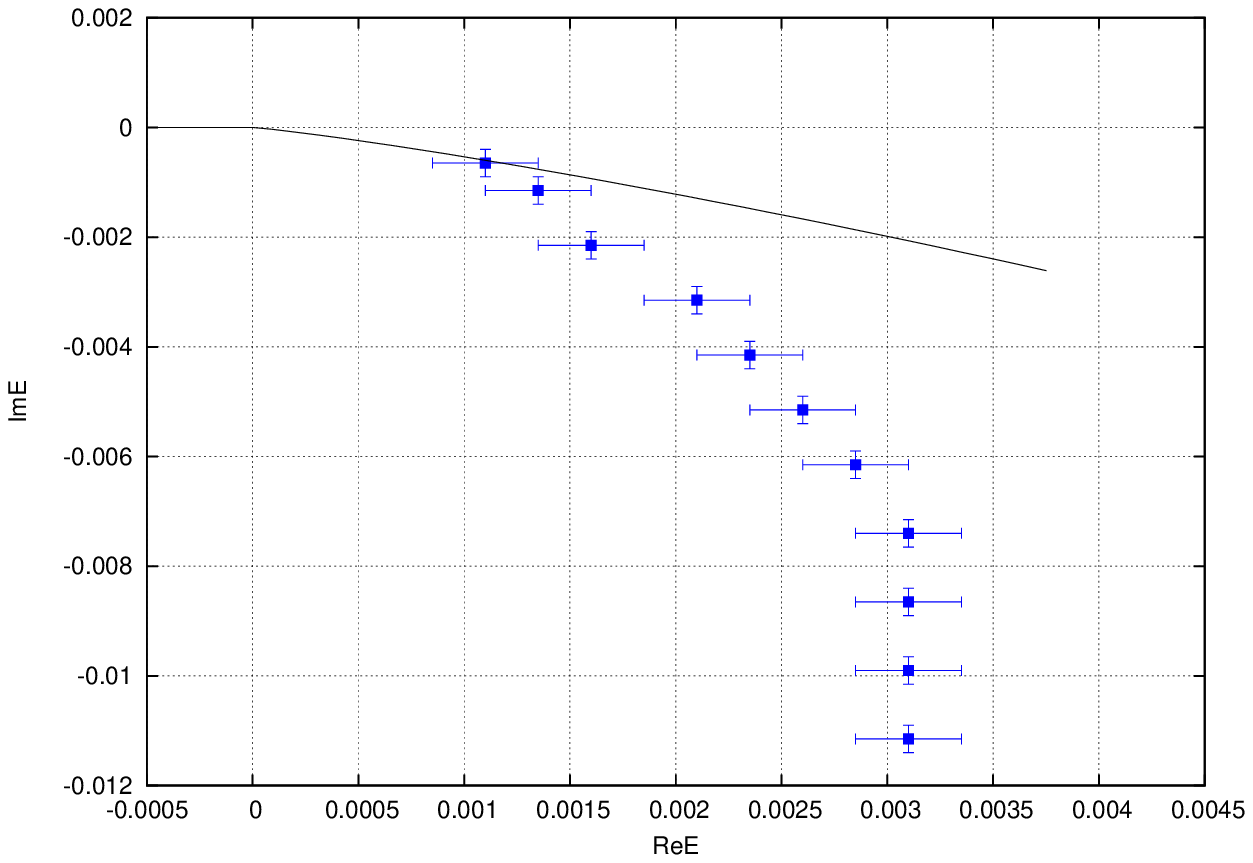}
		\end{minipage} \\
		\begin{minipage}[htbp]{0.5\hsize}
			\centering
			\includegraphics[width=8cm, angle = 0]{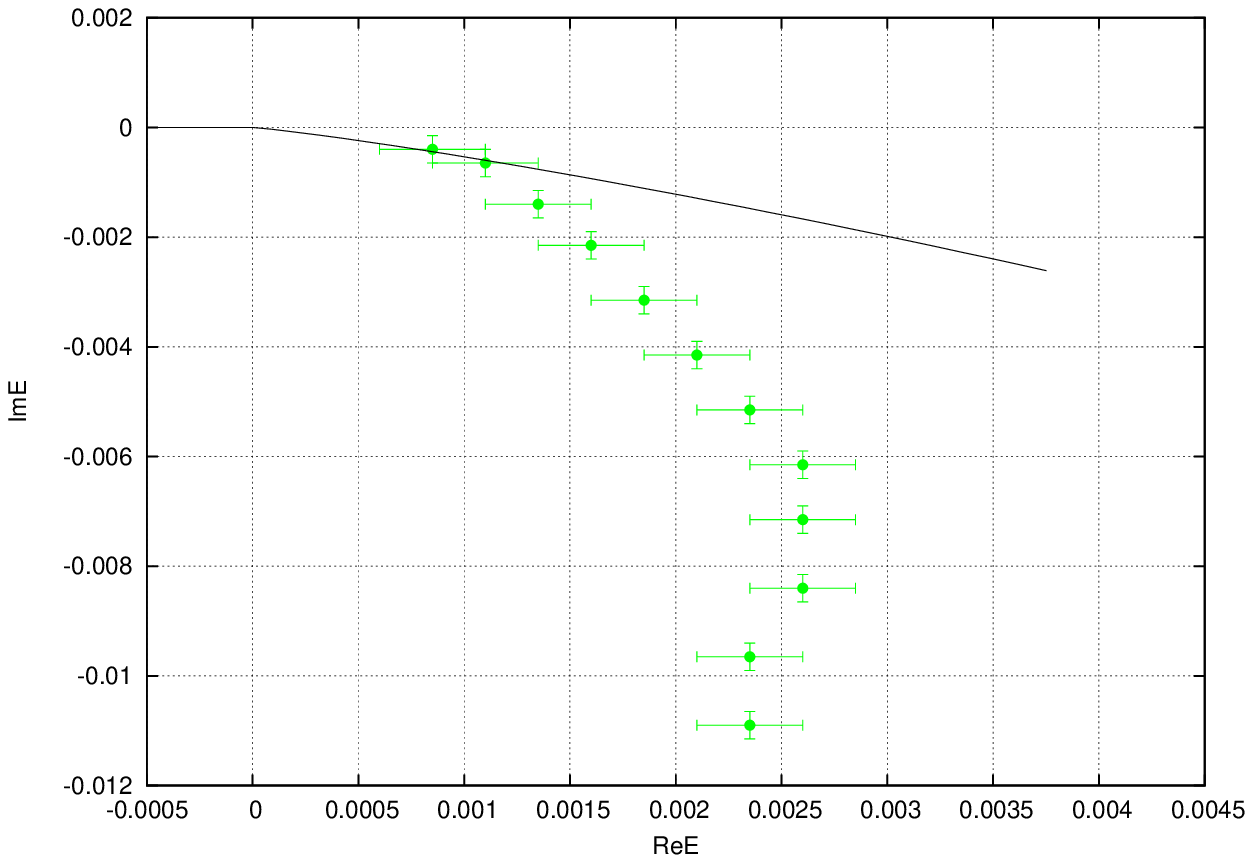}
		\end{minipage} &
		\begin{minipage}[htbp]{0.5\hsize}
			\centering
			\includegraphics[width=8cm, angle = 0]{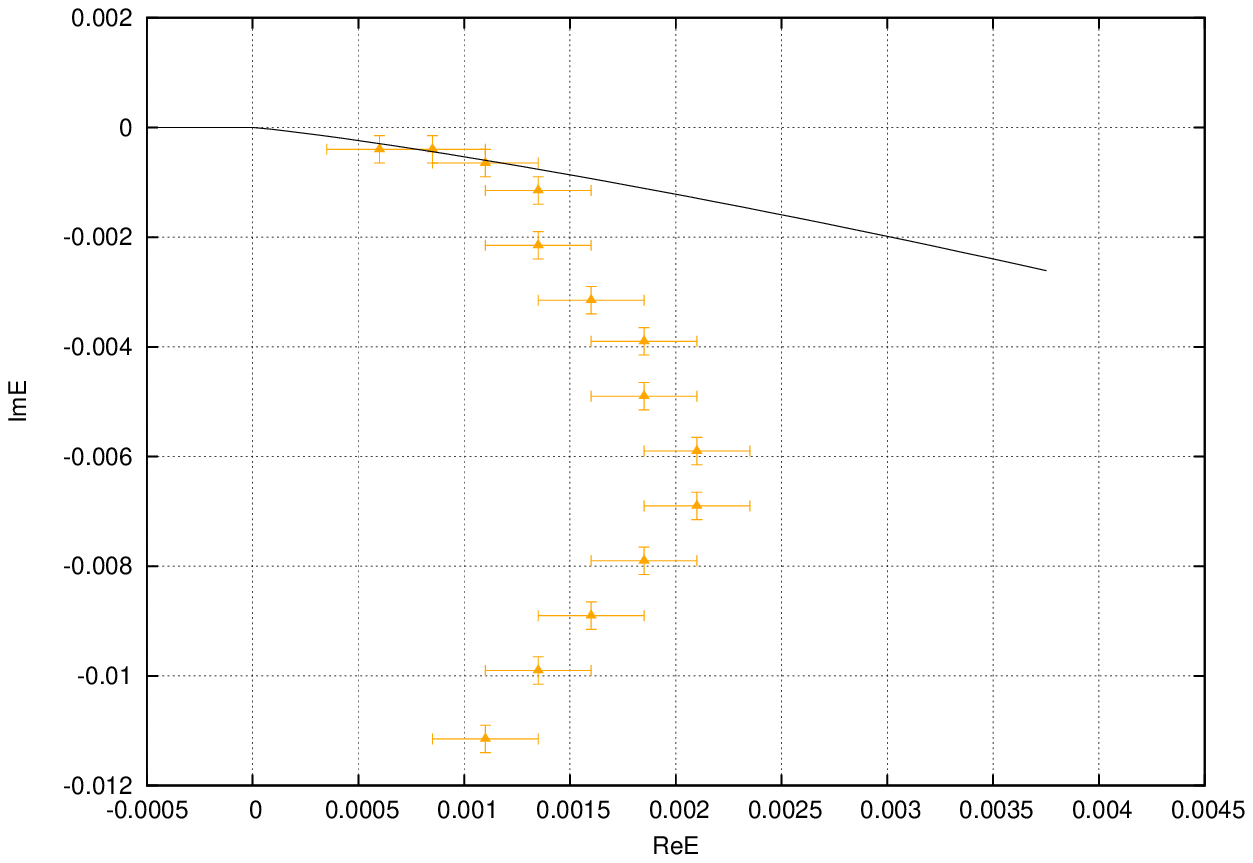}
		\end{minipage} \\
		\begin{minipage}[htbp]{0.5\hsize}
			\centering
			\includegraphics[width=8cm, angle = 0]{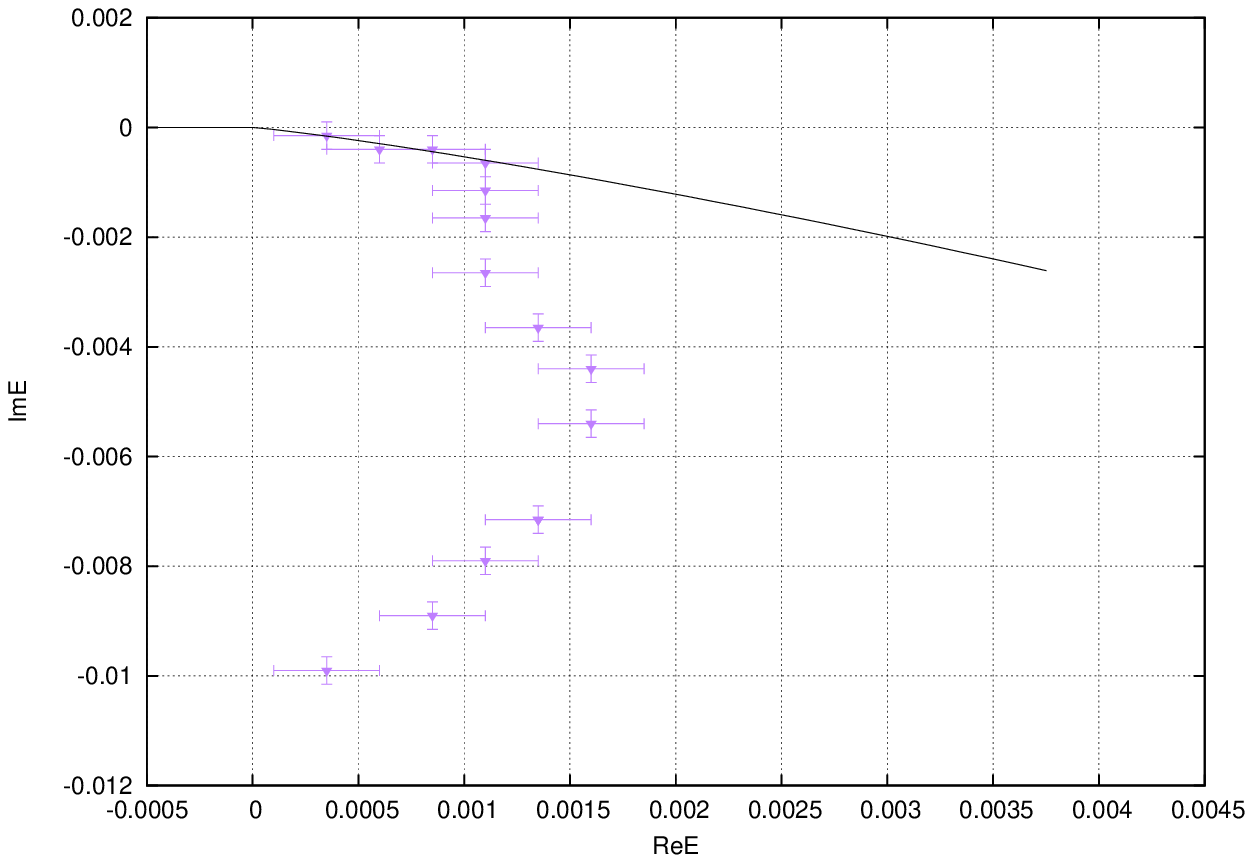}
		\end{minipage} &
		\begin{minipage}[htbp]{0.5\hsize}
			\centering
			\includegraphics[width=8cm, angle = 0]{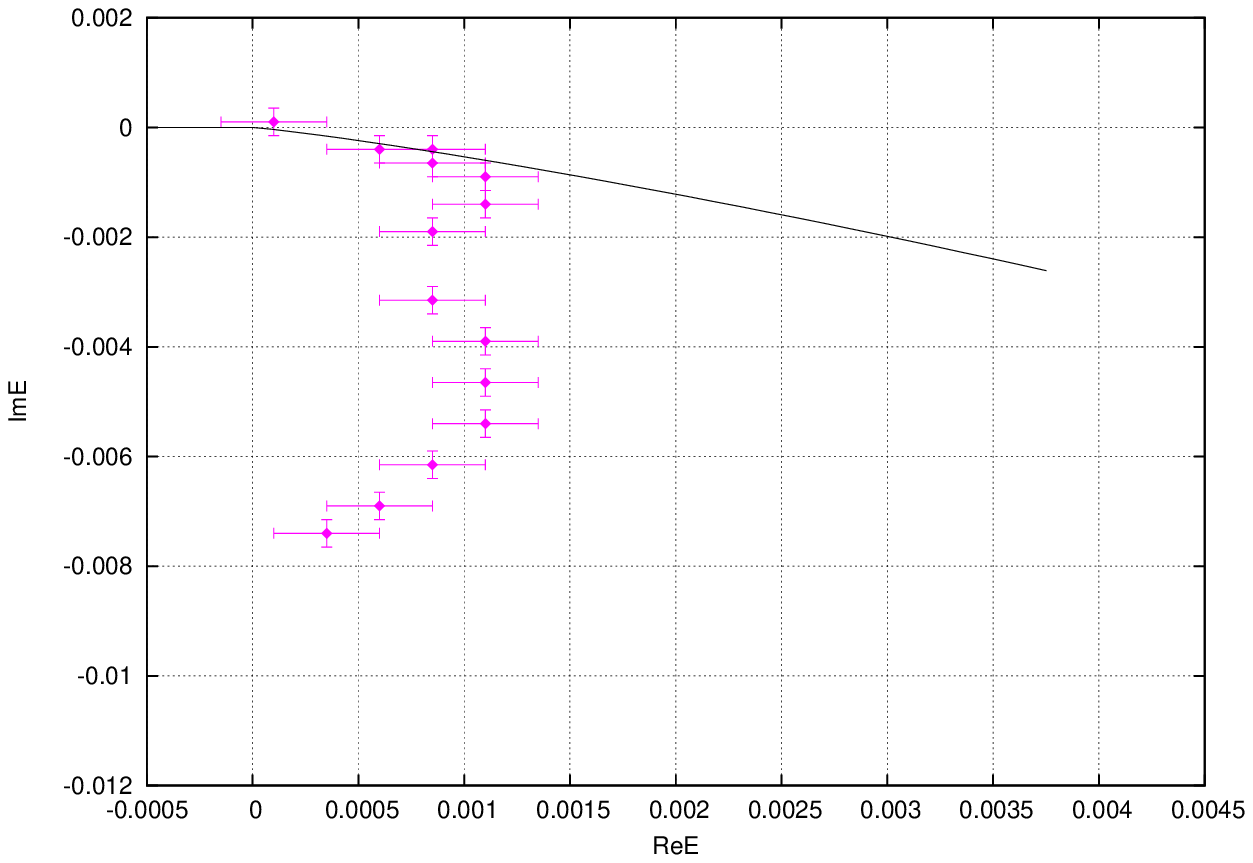}
		\end{minipage} \\
	\end{tabular}
	\caption{The $S$-matrix pole trajectories near the thresholds in a degenerate two-body and three-body coupled-channels system. Upper-left, upper-right, middle-left, middle-right, bottom-left and bottom-right figures correspond to $f_{\phi\phi} = - 0.15, - 0.20, - 0.25, - 0.30, - 0.35, - 0.40$. Each point in each figure corresponds to different $f_{\psi\phi_3}$. The solid curve is a trajectory determined by the equation $c + E \log{\left( - E \right)} = 0$, or equivallently ${\rm Im} \, E = \pi {\rm Re} \, E / \log \left({\rm Re} \, E\right)$. We can see that poles approach the thresholds from the fourth quadrant of the unphysical complex energy sheet. We can also see that when poles lie very close to the thresholds, they all lie on the solid curve irrespective of parameter sets.}
	\label{fig:poletrajectories1}
\end{figure}

\begin{figure}[htbp]
	\begin{tabular}{cc}
		\begin{minipage}[htbp]{0.5\hsize}
			\centering
			\includegraphics[width=8cm, angle = 0]{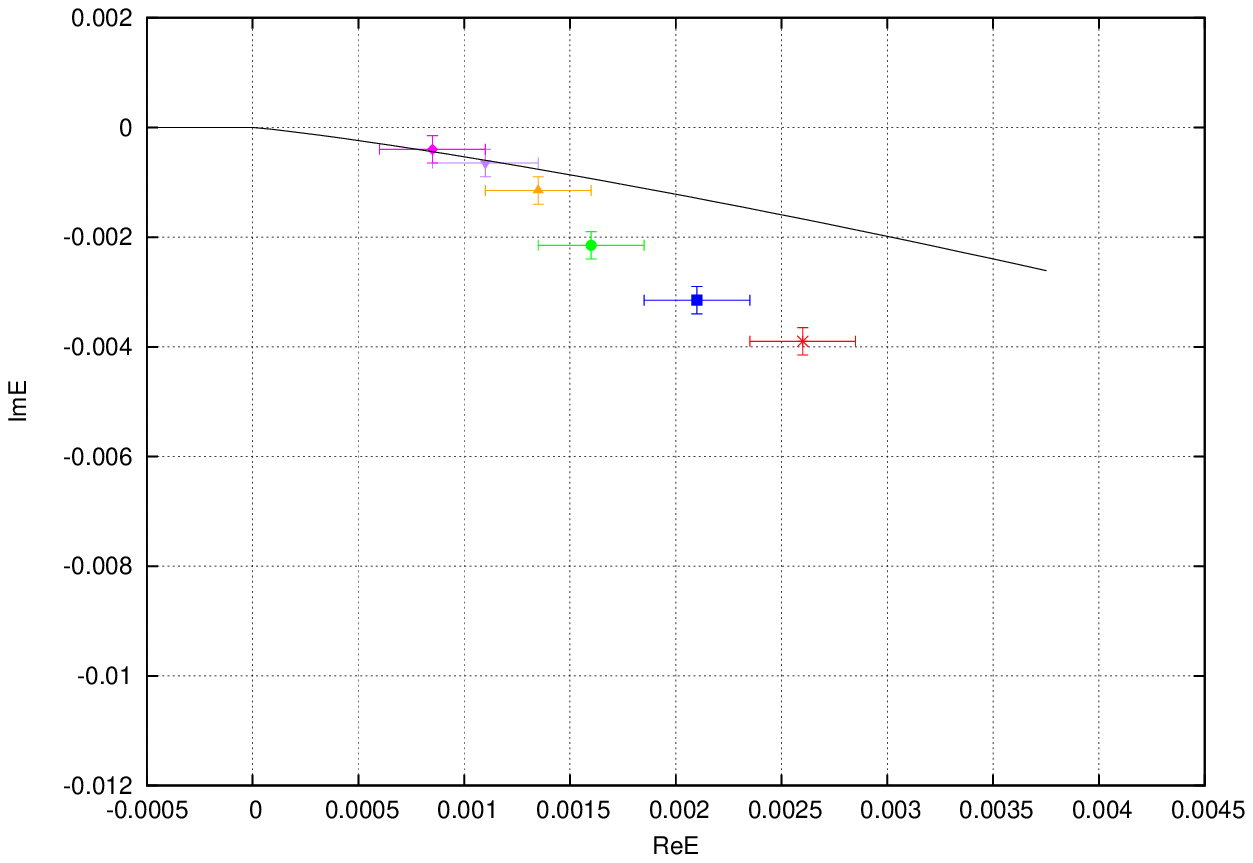}
		\end{minipage} &
		\begin{minipage}[htbp]{0.5\hsize}
			\centering
			\includegraphics[width=8cm, angle = 0]{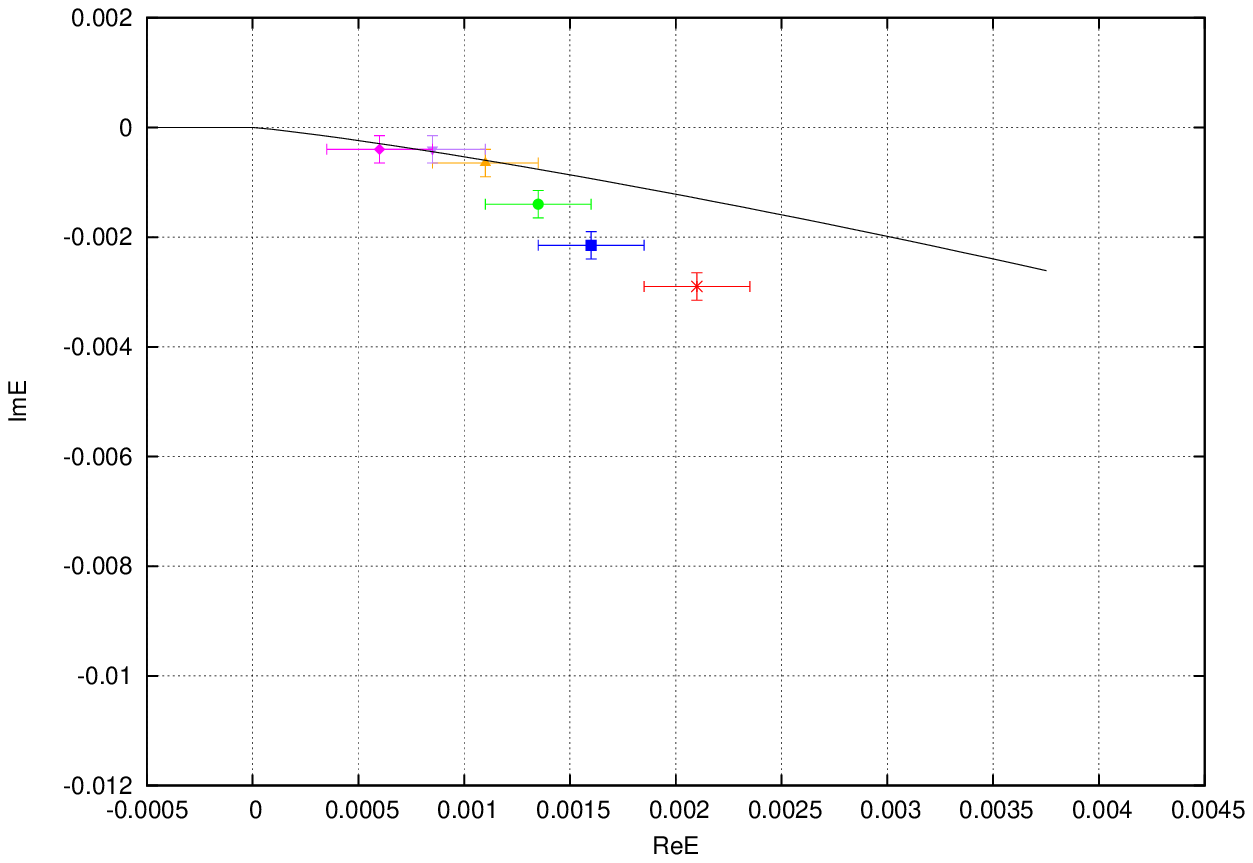}
		\end{minipage} \\
		\begin{minipage}[htbp]{0.5\hsize}
			\centering
			\includegraphics[width=8cm, angle = 0]{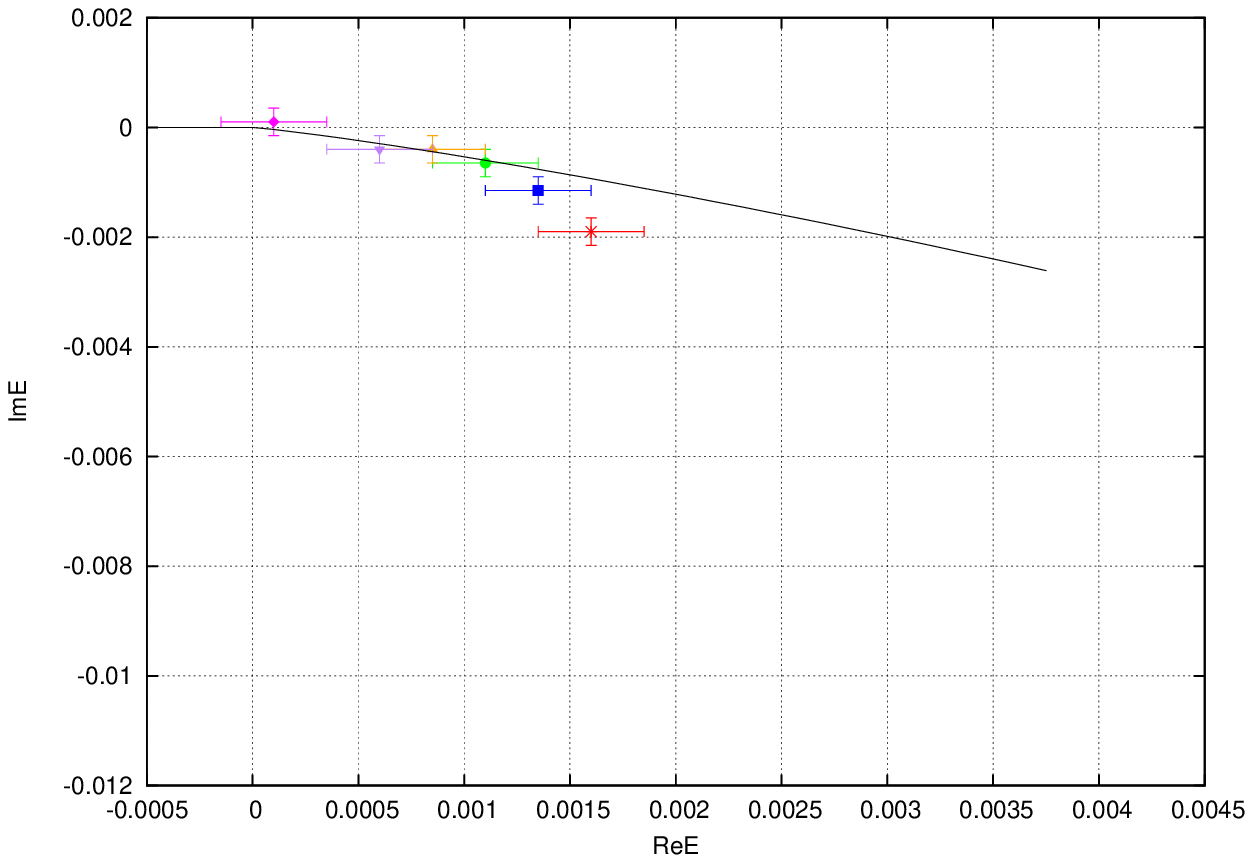}
		\end{minipage} &
	\end{tabular}
	\caption{The $S$-matrix pole behavior near the thresholds varying interactions in the three-body channel denoted as $V_{\phi_i\phi_j}$ while interaction in the two-body channel $V_{\psi\phi_3}$ is fixed. We can see that as the pole approaches the thresholds, they all lie on the curve determined by $c + E \log{\left( - E \right)} = 0$, or equivallently ${\rm Im} \, E = \pi {\rm Re} \, E / \log \left({\rm Re} \, E\right)$.}
	\label{fig:poletrajectories2}
\end{figure}

In Table \ref{table:poletrajectories}, we summarize obtained complex pole energies vs.\ coupling constants, $f_{\phi\phi}$ and $f_{\psi\phi}$.
Then, in Fig. \ref{fig:poletrajectories1} (Fig. \ref{fig:poletrajectories2}), we plot locations of poles in the unphysical complex energy sheet with error bars for different values of $f_{\psi\phi}$ ($f_{\phi\phi}$) with $f_{\phi\phi}$ ($f_{\psi\phi}$) fixed,
where we also show the curve determined by the equation,
\begin{equation}
	c + E \log{\left( - E \right)} = 0,
\end{equation}
or equivallently
\begin{equation}
	{\rm Im} \, E = \pi {\rm Re} \, E / \log \left({\rm Re} \, E\right).
\end{equation}

In Fig. \ref{fig:poletrajectories1} and Fig. \ref{fig:poletrajectories2} the trajectories of poles seem slightly unsmooth.
This might be due to the slow convergence of the integral equation when poles lie close to the rotated branch cut.
However, the deviations are still within errors.
 
We see that the $S$-matrix pole approaches the threshold from the fourth quadrant of the unphysical complex energy sheet as the absolute value of the coupling constant, $f_{\psi\phi}$ ($f_{\psi\psi}$), increases.
We also see that when poles lie close to the threshold, that is, when ${\rm Re} \, E \lesssim 0.001$, they all lie on the curve, $c + E \log{\left( - E \right)} = 0$.
This is consistent with the analytical study of Ref. \cite{KonishiMorimatsuYasui2}.
The $S$-matrix pole behavior near the threshold in the system is therefore universal in the sense that it is described by the unique one-parameter equation $c + E \log{\left( - E \right)} = 0$ irrespective of details of the specific parameter sets. \par

\begin{figure}%[htbp]
 \begin{minipage}{0.475\hsize}
  \begin{center}
   \includegraphics[width=3cm]{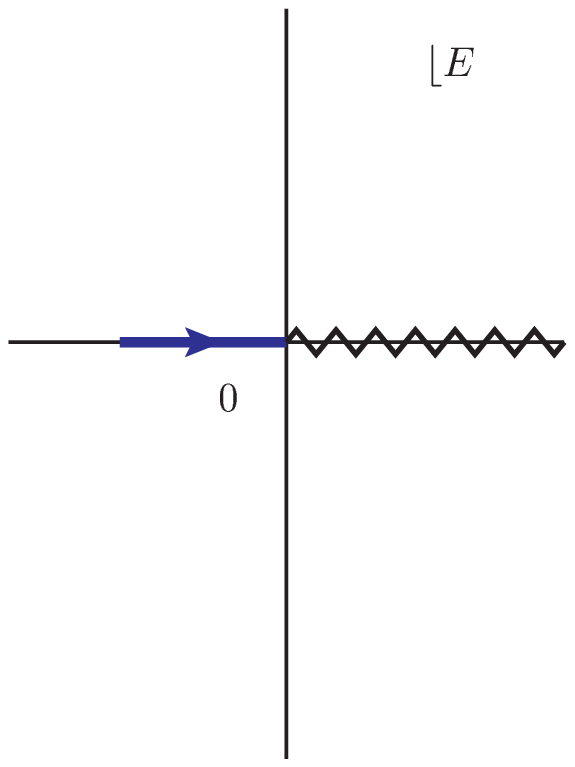}
  \end{center}
  \begin{center}
  The unphysical complex energy sheet.
  \end{center}
 \end{minipage}
 \begin{minipage}{0.475\hsize}
  \begin{center}
   \includegraphics[width=3cm]{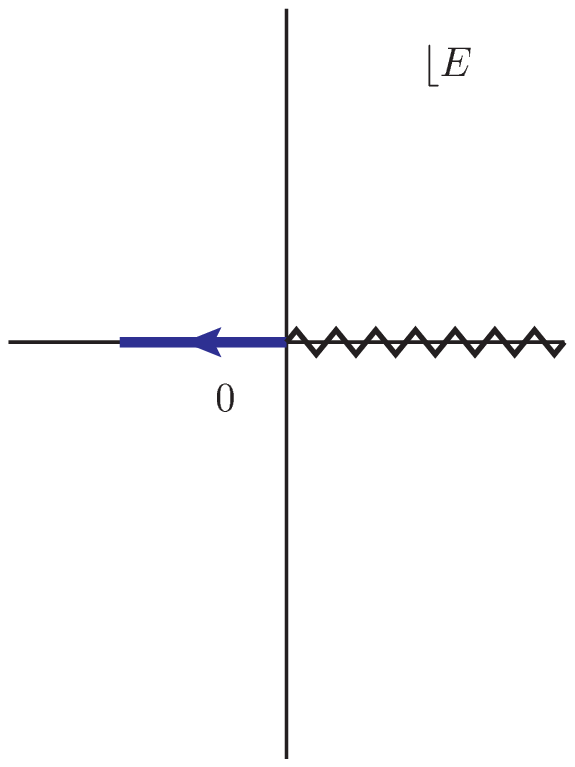}
  \end{center}
  \begin{center}
  The physical complex energy sheet.
  \end{center}
 \end{minipage}
 \caption{\label{} The $S$-matrix pole trajectory; $s$-wave.}
 \label{fig:pole1}
\end{figure}

\begin{figure}%[htbp]
 \begin{minipage}{0.475\hsize}
  \begin{center}
   \includegraphics[width=3cm]{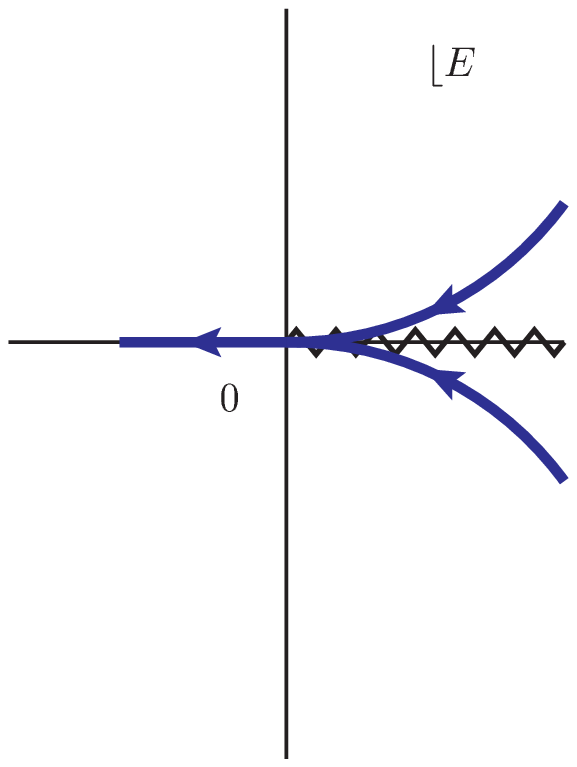}
  \end{center}
  \begin{center}
  The unphysical complex energy sheet.
  \end{center}
 \end{minipage}
 \begin{minipage}{0.475\hsize}
  \begin{center}
   \includegraphics[width=3cm]{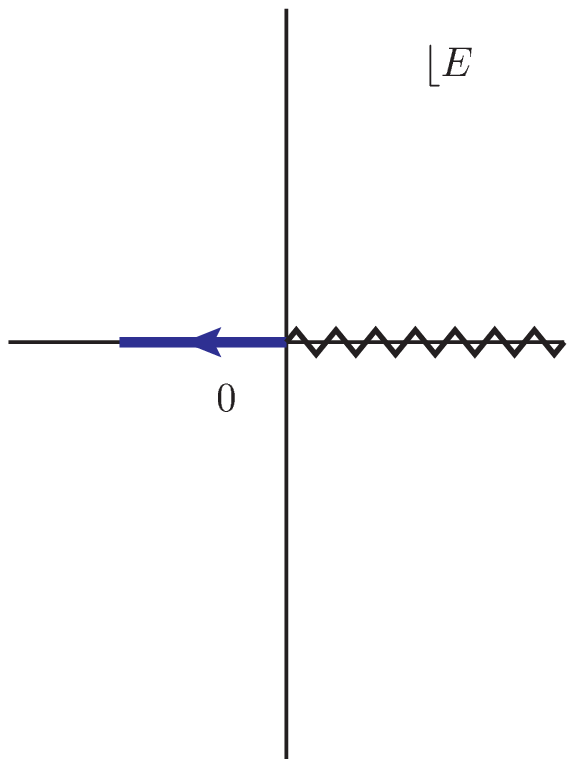}
  \end{center}
  \begin{center}
  The physical complex energy sheet.
  \end{center}
 \end{minipage}
 \caption{\label{} The $S$-matrix pole trajectory; higher-partial-wave.}
 \label{fig:pole2}
\end{figure}

%{\color{red} 
This behavior of the $S$-matrix pole near the threshold is different from either the two-body nor three-body scattering.
In the two-body scattering the $T$-matrix near the threshold is represented in the form of the effective range expansion and the pole is given as a solution of the equation
\begin{equation}\label{eq:ERE}
	- \frac{1}{a_{\ell}} + \frac{r_{\ell}}{2} p^2 + \cdots - i p^{ 2 \ell + 1 } = 0 \qquad \left(E = \frac{p^2}{2 \mu} \right),
\end{equation}
where $\ell$ is the relative angular momentum, $p$ and $\mu$ are the relative momentum and the reduced mass of the two particles, respectively.
From Eq.(\ref{eq:ERE}) the behavior of the pole momentum in the unphysical complex energy sheet is given as  
\begin{eqnarray}
	p &\sim& -i/a_0 \qquad (\ell = 0), \\
	p &\sim& \displaystyle{\pm\sqrt{\frac{2}{a_\ell r_\ell}} -i\frac{2^\ell}{a_\ell^\ell r_\ell^{\ell+1}}} .\qquad \left(\ell \ge 1 \right)
\end{eqnarray}
If $\ell=0$, the $S$-matrix pole approaches the threshold from the negative axis in the unphysical complex energy sheet and becomes a bound state as the interaction becomes more attractive.
If $\ell \ge 1$, the pole approaches the threshold from the fourth quadrant of the unphysical complex energy sheet, which manifests itself as a resonance if it lies close enough to the real axis, and becomes a bound state as the interaction becomes more attractive.
%The resonance becomes sharp as the relative angular momentum excitation is increased which can be understood that this is due to larger centrifugal barrier. 
Figs.\ref{fig:pole1} and \ref{fig:pole2} illustrate trajectories of poles of $\ell=0$ and $\ell \ge 1$, respectively, both in the physical and the unphysical complex energy sheet.
 
In the three-body scattering the $T$-matrix near the threshold behaves as in the two-body scattering with $\ell$ replaced by ${\cal L}=L+\frac{3}{2}$ where $L$ is the sum of two relative angular momenta in the three-body system \cite{Matsuyama:1991bm}.
Therefore, the behavior of the $S$-matrix pole in the three-body scattering is similar to that of the two-body scattering with $\ell \ge 1$ and resonances can exist irrespective of the relative angular momenta.


%{\color{red} The $S$-matrix pole behavior in a single-channel three-body system has also been investigated \cite{Matsuyama:1991bm}. In Ref \cite{Matsuyama:1991bm}, it is argued that a three-body system without relative angular momentum excitation corresponds to $\ell = 3/2$ two-body system. The $S$-matrix pole behavior near the threshold is therefore similar to that of two-body system with angular momentum excitation. Therefore, a resonance can exists even there are no relative angular momentum excitation in a single channel three-body system.}
\par
It should be noted that in the degenerate two-body and three-body coupled-channels system the $S$-matrix pole approaches from the fourth quadrant of the unphysical complex energy sheet even though the two-body system is in the $s$-wave, which is different from that of the two-body $s$-wave scattering.
Therefore, if poles lie close enough to the real axis, they might appear as resonances.

\section{Summary and Discussion}\label{summaryanddiscussion}
%\begin{itemize}
%\item
In this paper, we studied the $S$-matrix pole behavior near the threshold for the degenerate two-body and three-body coupled-channels system.
%\item
To that end, we formulated two-body and three-body coupled-channels scattering equations as effective three-body scattering equations,
effective AGS equations, by the Feshbach projection method.
In the effective AGS equations effects induced by the coupling to the two-body channel are embedded as effective interactions in the three-body channel.
%\item
%We call it the effective AGS equations and the effective interactions in the three-body channel are constructed by the Feshbach projection method. 
%\item
Even in the absence of elementary three-body interactions, the coupling to the two-body channel generates effective three-body interactions. 
%\item

We solved the eigenvalue equation of the kernel of the scattering equations instead of solving the equations themselves to obtain the $S$-matrix pole energy. 
%\item
However, we faced the problem of an unphysical singularity when we na{\' i}vely tried to solve the eigenvalue equation.
%\item
Namely, it turns out that the solution of the eigenvalue equation has an unphysical branch point due to the fact that the physical mass of a particle is different from the bare one.
%\item
The full transition amplitudes of course have only the physical singularities, i.e.\ it has a physical branch point with the physical mass but not a unphysical branch point with the bare mass.
%\item
We showed that this problem is resolved by an appropriate reorganization of the scattering equations and the mass renormalization.
%\item
In a word, we included the counterterms which appear in higher order terms of the scattering equations into the kernel of the scattering equations, which is an input of the eigenvalue equation.
%\item

We numerically solved the modified eigenvalue equation and obtained the $S$-matrix pole behavior near the thresholds for various parameters.
%\item
The $S$-matrix pole approaches the threshold from the fourth quadrant of the unphysical energy sheet, which manifests itself as a resonance if it lies close enough to the real axis, and becomes a bound state as the interaction becomes more attractive.
%\item
The behavior of the obtained numerical results is consistent with the universal behavior $c + E \log{\left( - E \right)} = 0$ analytically obtained in Ref.\cite{KonishiMorimatsuYasui2}
%\item
This behavior is different from the two-body or three-body scatterings and is characteristic in the degenerate two-body and three-body coupled-channels system.
%This behavior is in contrast to degenerate coupled-channels two-body system in which one of the channel is in $s$-wave.
%\item
This characteristic behavior may be observed in the cold atom system in which the interaction can be controlled by applying external fields or (approximately) in the hadronic systems where two-body and three-body thresholds are (approximately) degenerate.
%}
%\end{itemize}

\begin{acknowledgments}
We would like to thank Pascal Naidon for discussions about Efimov physics in general and also useful comments on our work.
We would also like to thank Hiroyuki Kamano, Toru Sato and Koichi Yazaki for helpful discussions.
We would also like to thank computational environment at the High Energy Accelerator Research Organization (KEK).
The work of Atsunari Konishi is partly supported by research assistant for the university of Tokyo grants for PhD research and research assistant grants by the High Energy Accelerator Research Organization (KEK).
\end{acknowledgments}

\pagebreak

\appendix

\section{Modified Kernel in the presence of $V_{\phi_1\phi_2}$}
\label{sec:appendix1}
In section \ref{setup}, we derived the modified kernel by reorganizing the Feynman diagrams order-by-order with respective to $V_{\psi\phi_3}$, the interaction between $\psi\phi_3$. There, we ignored the elementary interaction between $\phi_1\phi_2$ denoted as $V_{\phi_1\phi_2}$ to simplify the argument. In this appendix, we present a detailed derivation of the modified kernel taking also $V_{\phi_1\phi_2}$ into account and by explicitly summing each terms in the effective AGS equations as a geometric series. In the following, we suppress the argument of energy $E$ for notational simplicity.

\subsection{The self-energies in higher-order terms in the effective AGS equations}
In this subsection, we see how the self-energies appear in higher-order terms of the effective AGS equations and how the corresponding counterterms are added to it. Derivation of the modified kernel is given in the next section. \par
In the presence of $V_{\phi_1\phi_2}$, the self-energy can be decomposed of two parts, one that contains $\phi_1\phi_2$ interaction which we denote as $\Sigma_V$ and the other does not denoted as $\Sigma_0$
\begin{equation}
	\Sigma = \Sigma_0 + \Sigma_V.
\end{equation}
Its diagrammatic representation is given as
\begin{equation}
	\raisebox{-0.3cm}{\includegraphics[width=6cm]{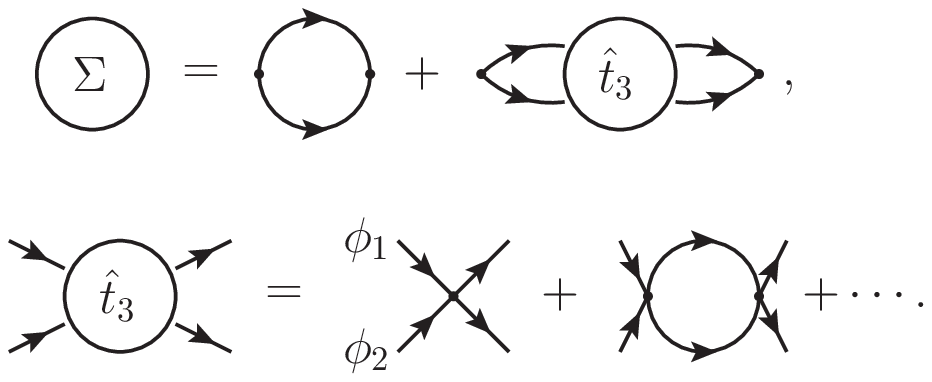}}
\end{equation}
When it comes to consider the self-energies which appear in higher-order terms of the effective AGS equations, it is convenient to write $t_3$ in the following form
\begin{equation}
	t_3 = {\hat t_3} + \left( 1 +{\hat t_3} G_0^{\phi\phi\phi} \right) V_{\psi\mathchar`-\phi_1\phi_2} G^{\psi\phi_3} V_{\phi_1\phi_2\mathchar`-\psi} \left( G_0^{\phi\phi\phi} {\hat t_3} + 1 \right),
\end{equation}
where ${\hat t_3} \left( E \right)$ is defined by
\begin{equation}
	{\hat t_3} = V_3 \frac{1}{1 - G_0^{\phi\phi\phi} V_3}.
\end{equation}
Diagrammatic representation of $t_3$ is given as %in Fig. \ref{fig:t3}.
\begin{equation}
	\includegraphics[width=17cm]{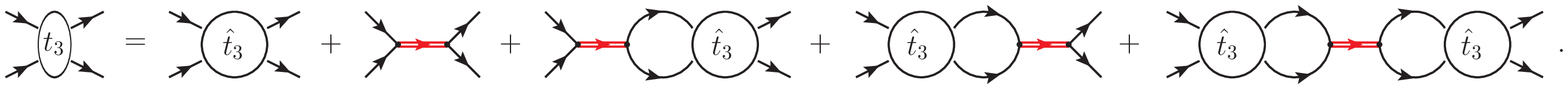}
\end{equation}
%\begin{figure}[htbp]
%	\includegraphics[width=17cm]{t3_1.eps}
%	\caption{A diagrammatic representation of $t_{\phi_1\phi_2}$.}
%	\label{fig:t3}
%\end{figure}
\par
For notational simplicity, we rewrite $Z_4$ as follows
\begin{equation}
	Z_4 = z_4 {\bf 1}, \hspace{0.5cm} z_4 = G_0^{\phi\phi\phi} V_{\psi\mathchar`-\phi_1\phi_2} {\cal G} V_{\phi_1\phi_2\mathchar`-\psi} G_0^{\phi\phi\phi},
	\label{eq:z4}
\end{equation}
where we defined ${\cal G}$ by
\begin{equation}
	{\cal G} = G_0^{\psi\phi_3} t_{\psi\phi_3} \frac{1}{1 - G_0^{\psi\phi_3} \Sigma_0 t_{\psi\phi_3}} G_0^{\psi\phi_3},
	\label{eq:G}
\end{equation}
and $t_{\psi\phi_3}$ by
\begin{equation}
	t_{\psi\phi_3} = V_{\psi\phi_3} + V_{\psi\phi_3} \frac{1}{E - H_0^Q - V_{\psi\phi_3}} V_{\psi\phi_3},
\end{equation}
whose diagrammatic representation is,
\begin{equation}
		\raisebox{0cm}{\includegraphics[width=7.5cm]{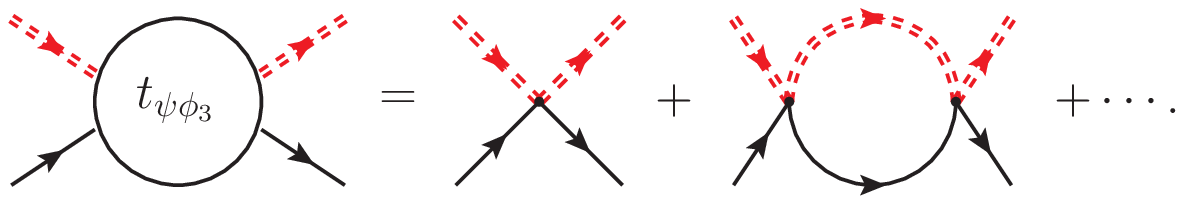}}
\end{equation}
\par
$z_4 t_3 z_4$, which appears in higher order terms of the effective AGS equations, is written as
\begin{eqnarray}
	z_4 t_3 z_4 & = & G_0^{\phi\phi\phi} V_{\psi\mathchar`-\phi_1\phi_2} {\cal G} \left( \Sigma_V + \Sigma G^{\psi\phi_3} \Sigma \right) {\cal G} V_{\psi\mathchar`-\phi_1\phi_2} G_0^{\phi\phi\phi} \nonumber \\
	& = & \raisebox{-0.45cm}{\includegraphics[width=13cm]{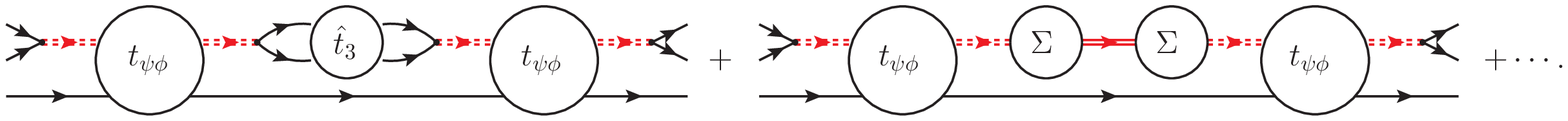}}
	\label{eq:z4t3z4}
\end{eqnarray}
The corresponding $\Sigma_0$ for the first term in the above is actually included in the second term of $z_4$ as
\begin{equation}
	\raisebox{-0.4cm}{\includegraphics[width=11.5cm]{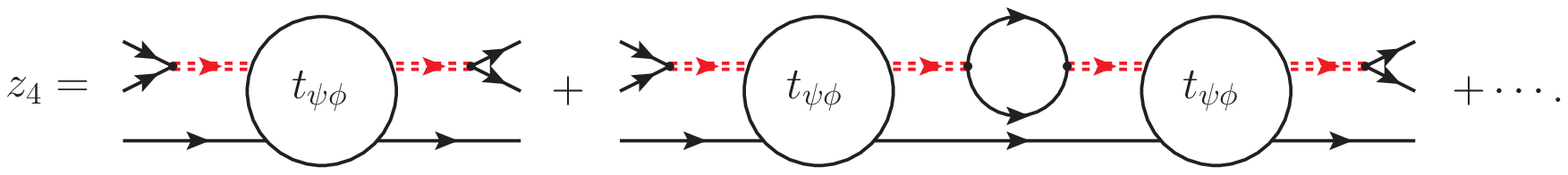}}
\end{equation}
Combining the second term of the above equation with the first term of Eq. (\ref{eq:z4t3z4}), we obtain
\begin{eqnarray}
	& & G_0^{\phi\phi\phi} V_{\psi\mathchar`-\phi_1\phi_2} G_0^{\psi\phi_3} t_{\psi\phi_3} G_0^{\psi\phi_3} \left( \Sigma + \Sigma G^{\psi\phi_3} \Sigma \right) G_0^{\psi\phi_3}  t_{\psi\phi_3} G_0^{\psi\phi_3} V_{\phi_1\phi_2\mathchar`-\psi} G_0^{\phi\phi\phi} \nonumber \\
	& = & \raisebox{-0.5cm}{\includegraphics[width=13cm]{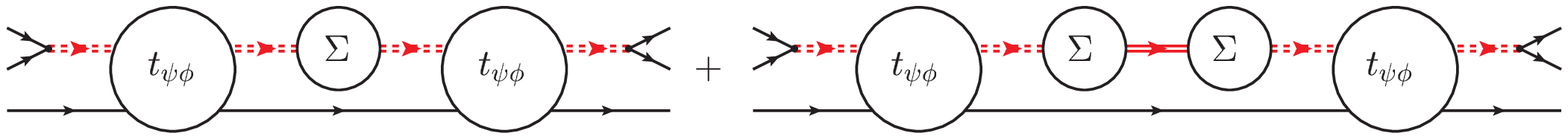}}.
\end{eqnarray}
The corresponding counterterms are as follows
\begin{eqnarray}
	& & G_0^{\phi\phi\phi} V_{\psi\mathchar`-\phi_1\phi_2} t_{\psi\phi_3} \left( \Delta + \Delta G^{\psi\phi_3} \Sigma + \Sigma G^{\psi\phi_3} \Delta + \Delta G^{\psi\phi_3} \Delta \right) t_{\psi\phi_3} V_{\psi\mathchar`-\phi_1\phi_2} G_0^{\phi\phi\phi} \nonumber \\
	& & \raisebox{-0.0cm}{\includegraphics[width=7.5cm]{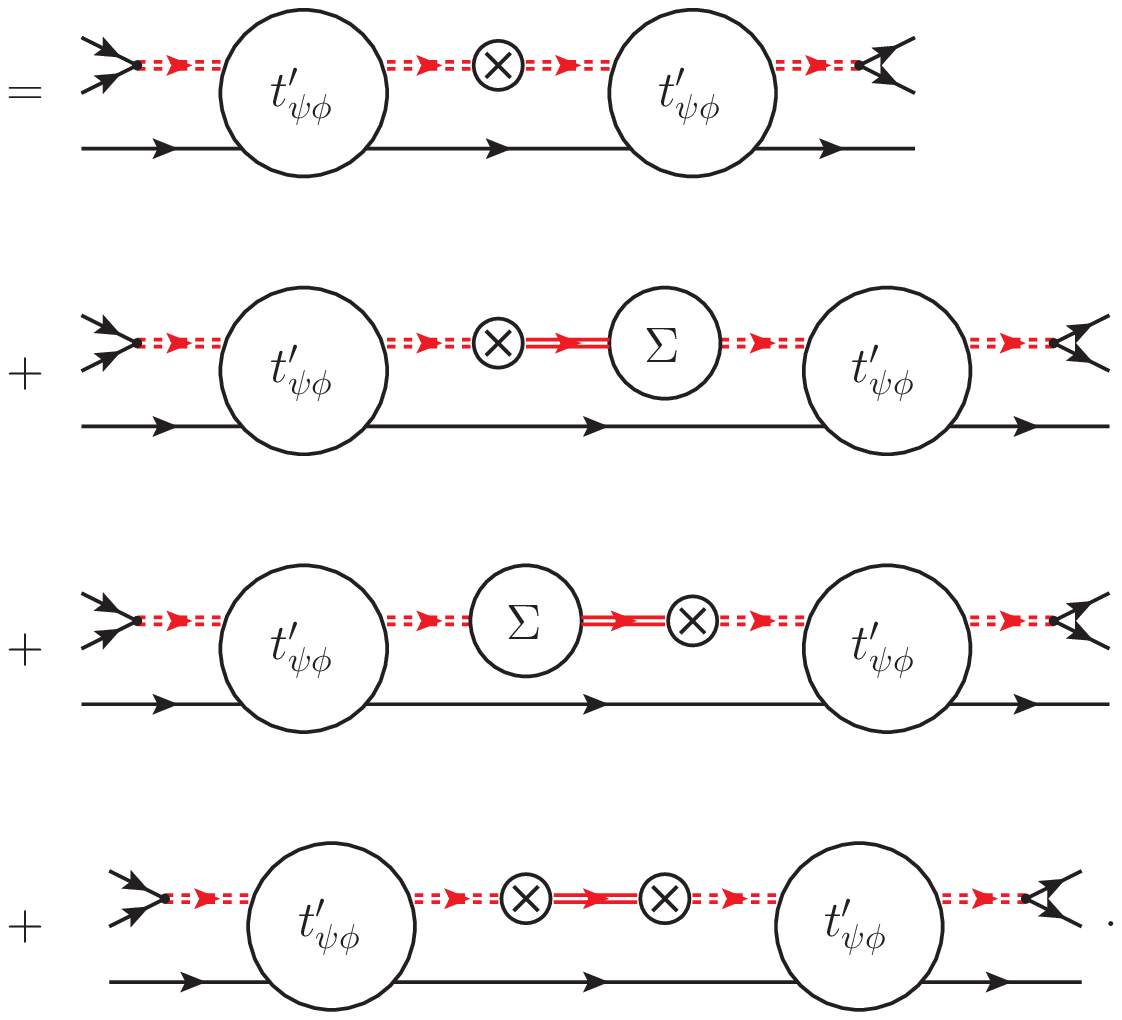}}
\end{eqnarray}
\par
Another way the self-energies appear in higher order terms of the effective AGS equations is ones like $ Z_0 T Z_4 $ which contains the matrix elements  $ G_0^{\phi\phi\phi} t_3 z_4 $ which is rewritten as
\begin{eqnarray}
	G_0^{\phi\phi\phi} t_3 z_4 & = & G_0^{\phi\phi\phi} \left( {\hat t_3} + \left( 1 + {\hat t_3} G_0^{\phi\phi\phi} \right) V_{\psi\mathchar`-\phi_1\phi_2} G^{\psi\phi_3} V_{\phi_1\phi_2\mathchar`-\psi} \left( G_0^{\phi\phi\phi} {\hat t_3} + 1 \right) \right) G_0^{\phi\phi\phi} V_{\psi\mathchar`-\phi_1\phi_2} {\cal G} V_{\psi\mathchar`-\phi_1\phi_2} G_0^{\phi\phi\phi} \nonumber \\
	& = & G_0^{\phi\phi\phi} \left( {\hat t_3} G_0^{\phi\phi\phi} V_{\psi\mathchar`-\phi_1\phi_2} + \left( 1 + {\hat t_3} G_0^{\phi\phi\phi} \right) V_{\psi\mathchar`-\phi_1\phi_2} G^{\psi\phi_3} \Sigma \right) {\cal G} V_{\psi\mathchar`-\phi_1\phi_2} G_0^{\phi\phi\phi},
\end{eqnarray}
whose diagrammatic representation is given below
\begin{equation}
	\raisebox{-0.3cm}{\includegraphics[width=13cm]{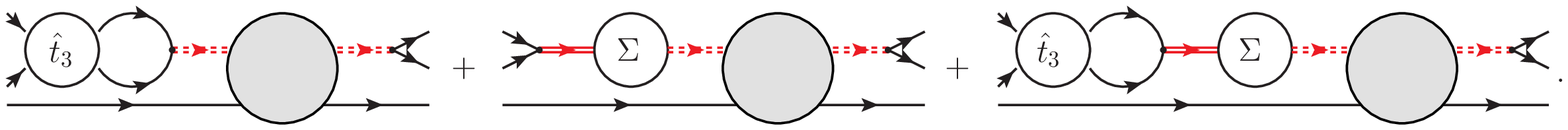}}
\end{equation}
The corresponding counterterms are therefore,
\begin{widetext}
\begin{equation}
	G_0^{\phi\phi\phi} \left( 1 + {\hat t_3} G_0^{\phi\phi\phi} \right) V_{\psi\mathchar`-\phi_1\phi_2} G^{\psi\phi_3} \Delta\ {\cal G} V_{\psi\mathchar`-\phi_1\phi_2} G_0^{\phi\phi\phi}
	= \raisebox{-0.45cm}{ \includegraphics[width=8cm]{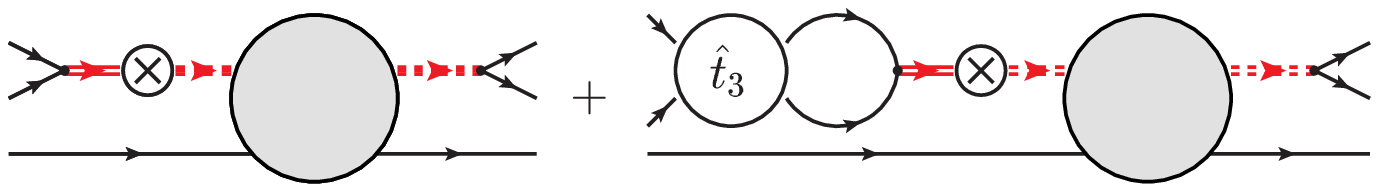} }.
\end{equation}
\end{widetext}
Having seen what sort of the self-energies appear in higher-order terms of the effective AGS equations, we will find a way to put those counterterms into one of the driving term $Z_4$ and define the modified one in the next subsection.

\subsection{Reorganization of the Effective AGS equations and the modified kernel}
In the following, we perform the mass renormalization and reorganize scattering processes in the effective AGS equations so as to keep the structure of the equations the same while the counterterms in higher-order terms of the equations are included into the kernel of the equations. \par
The self-energies appearing in higher-order terms of the effective AGS equations are generated when one of the driving term, $Z_4$, and one of the two-body $T$-matrix $t_3$ are multiplied. We therefore sum up $t_3$ first so that the additional self-energy does not appear in higher-order terms as discussed in detail below. 
\par
%Discussion about the mass renormalization here.
%\par
As we saw in section \ref{setup}, the transition amplitude $X$ is reorganized as
\begin{equation}
	X = \frac{1}{1 - Z T} Z = X_3 \frac{1}{1 - {\bar T_3} X_3}.
\end{equation}
Noting that $Z_0 T_3 Z_0 = 0$, we can simplify $X_3$ as
\begin{equation}
	X_3 = Z_0 + Z_0 T_3 Z_0 + \left( 1 + Z_0 T_3 \right) W_3 \left( T_3 Z_0 + 1 \right),
\end{equation}
where we defined $W_3$ by
\begin{equation}
	W_3 = Z_4 \frac{1}{1 - T_3 Z_4}.
\end{equation}
We note that the additional self-energy does not appear when ${\bar T_3}$ and $X_3$ are multiplied. $W_3$ is channel independent $W_3 = w_3 {\bf 1}$, where $w_3$ is defined by
\begin{equation}
	w_3 = z_4 + z_4 t_3 z_4 + \cdots = z_4 \frac{1}{1 - t_3 z_4}.
\end{equation}
$w_3$ can be expressed in some ways. Substituting an expression in Eq. (\ref{eq:z4}), $w_3$ is written as
\begin{eqnarray}
	w_3 & = & G_0^{\phi\phi\phi} V_{\psi\mathchar`-\phi_1\phi_2} {\cal G} \frac{1}{1 - V_{\phi_1\phi_2\mathchar`-\psi} G_0^{\phi\phi\phi} t_3 G_0^{\phi\phi\phi} V_{\psi\mathchar`-\phi_1\phi_2} {\cal G}} V_{\phi_1\phi_2\mathchar`-\psi} G_0^{\phi\phi\phi} \nonumber \\
	& = & G_0^{\phi\phi\phi} V_{\psi\mathchar`-\phi_1\phi_2} {\cal G} \frac{1}{1 - \left( \Sigma_V + \Sigma G^{\psi\phi_3} \Sigma \right) {\cal G}} V_{\phi_1\phi_2\mathchar`-\psi} G_0^{\phi\phi\phi} \nonumber \\
	& = & \raisebox{-0.475cm}{\includegraphics[width=15cm]{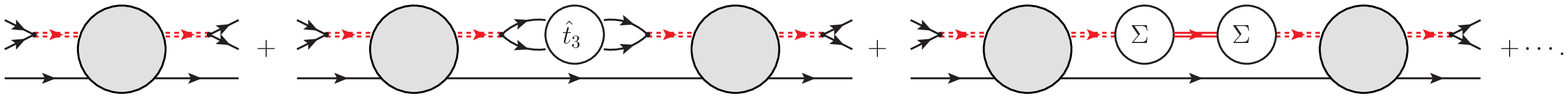}}
\end{eqnarray}
Substituting the definition of ${\cal G}$ Eq.(\ref{eq:G}), we obtain the following expression
\begin{eqnarray}
	w_3 & = & G_0^{\phi\phi\phi} V_{\psi\mathchar`-\phi_1\phi_2} G_0^{\psi\phi_3} t_{\psi\phi_3} G_0^{\psi\phi_3} \frac{1}{1 - \left( \Sigma + \Sigma G^{\psi\phi_3} \Sigma \right) G_0^{\psi\phi_3} t_{\psi\phi_3} G_0^{\left( 2 \right)}} V_{\phi_1\phi_2\mathchar`-\psi} G_0^{\phi\phi\phi} \nonumber \\
	& = & \raisebox{-0.475cm}{\includegraphics[width=15cm]{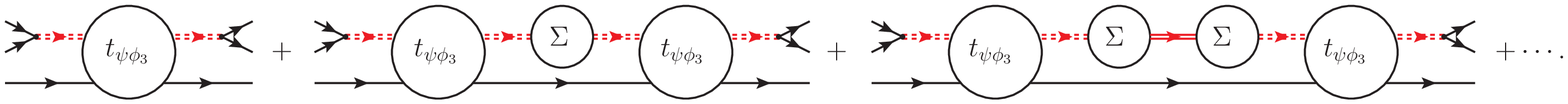}}
\end{eqnarray}
Noting that
\begin{equation}
	G_0^{\psi\phi_3} \left( \Sigma + \Sigma G^{\psi\phi_3} \Sigma \right) G_0^{\psi\phi_3} = G^{\psi\phi_3}  - G_0^{\psi\phi_3},
\end{equation}
we obtain another expression
\begin{eqnarray}
	w_3 & = & G_0^{\phi\phi\phi} V_{\psi\mathchar`-\phi_1\phi_2} G_0^{\psi\phi_3} t_{\psi\phi_3} \frac{1}{1 - \left( G^{\psi\phi_3} - G_0^{\psi\phi_3} \right) t_{\psi\phi_3} } G_0^{\psi\phi_3} V_{\phi_1\phi_2\mathchar`-\psi} G_0^{\phi\phi\phi} \nonumber \\
	& & \includegraphics[width=17cm]{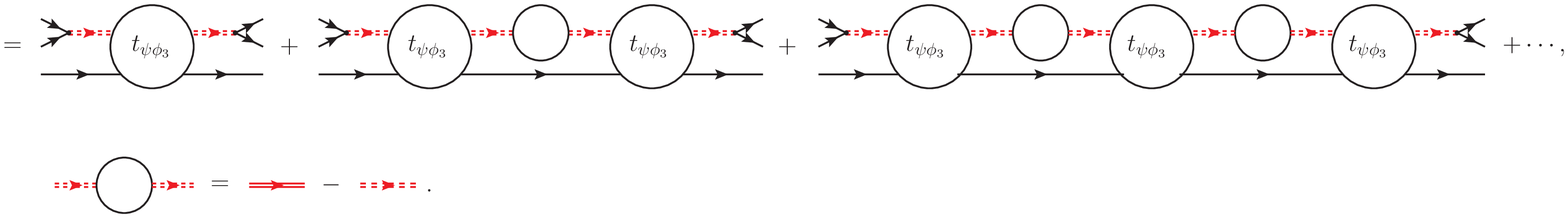}
\end{eqnarray}
Substituting the definition of $t_{\psi\phi_3}$, we have the most intuitive expression of $w_3$
\begin{eqnarray}
	w_3 & = & G_0^{\phi\phi\phi} V_{\psi\mathchar`-\phi_1\phi_2} G_0^{\psi\phi_3} V_{\psi\phi_3} \frac{1}{1 - G^{\psi\phi_3} V_{\psi\phi_3}} G_0^{\psi\phi_3} V_{\phi_1\phi_2\mathchar`-\psi} G_0^{\phi\phi\phi} \nonumber \\
	& = & G_0^{\phi\phi\phi} V_{\psi\mathchar`-\phi_1\phi_2} G_0^{\psi\phi_3} t_{\psi\phi_3}^{\Sigma} G_0^{\psi\phi_3} V_{\phi_1\phi_2\mathchar`-\psi} G_0^{\phi\phi\phi} \nonumber \\
	& = & \raisebox{-0.6cm}{ \includegraphics[width=8cm]{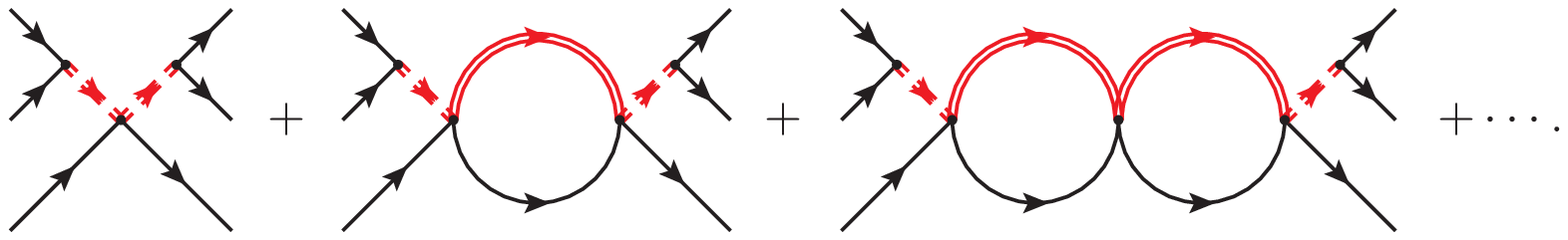} }
\end{eqnarray}
The intermediate factor $t_{\psi\phi_3}^{\Sigma} = V_{\psi\phi_3} / \left( 1 - G^{\psi\phi_3} V_{\psi\phi_3} \right)$ is represented in terms of the renormalized quantities as follows
\begin{eqnarray}
	t_{\psi\phi_3}^{\Sigma} & = & t_{\psi\phi_3}' + t_{\psi\phi_3}' \left( G^{\psi\phi_3} - G_0^{'\psi\phi_3} \right) t_{\psi\phi_3}' + \cdots \nonumber \\
	& = & t_{\psi\phi_3}' + t_{\psi\phi_3}' G_0^{'\psi\phi_3} \left( \Sigma + \Delta + \left( \Sigma + \Delta \right) G^{\psi\phi_3} \left( \Sigma + \Delta \right) \right) G_0^{'\psi\phi_3} t_{\psi\phi_3}' + \cdots,
\end{eqnarray}
where we introduced the renormalized free Green function
\begin{equation}
	G_0^{'\psi\phi_3} = \left( E - M' - m_3 - \frac{p_3^2}{2 M'} - \frac{k_3^2}{2 m_3} \right)^{-1},
\end{equation}
and another two-body $T$-matrix
\begin{equation}
	t_{\psi\phi_3}' = V_{\psi\phi_3} + V_{\psi\phi_3} G_0^{'\psi\phi_3} V_{\psi\phi_3} + \cdots.
\end{equation}
Its Green function is the bare one instead of the dressed one. $w_3$ is therefore written in terms of the renormalized quantities as
\begin{eqnarray}
	w_3 & = & G_0^{\phi\phi\phi} V_{\psi\mathchar`-\phi_1\phi_2} G_0^{\psi\phi_3} t_{\psi\phi_3}' \frac{1}{1 - \left( G^{\psi\phi_3} - G_0^{'\psi\phi_3} \right) t_{\psi\phi_3}'} G_0^{\psi\phi_3} V_{\phi_1\phi_2\mathchar`-\psi} G_0^{\phi\phi\phi} \nonumber \\
	& = & G_0^{\phi\phi\phi} V_{\psi\mathchar`-\phi_1\phi_2} G_0^{\psi\phi_3} t_{\psi\phi_3}' \frac{1}{1 - G_0^{'\psi\phi_3} \left( \Sigma + \Delta + \left( \Sigma + \Delta \right) G^{\psi\phi_3} \left( \Sigma + \Delta \right) \right) G_0^{'\psi\phi_3} t_{\psi\phi_3}'} G_0^{\psi\phi_3} V_{\phi_1\phi_2\mathchar`-\psi} G_0^{\phi\phi\phi} \nonumber \\
	& = & G_0^{\phi\phi\phi} V_{\psi\mathchar`-\phi_1\phi_2} {\cal G}' \frac{1}{1 - \left( \Sigma_V + \Sigma G_{\Sigma}^{\psi\phi_3} \Sigma \right) {\cal G}'} V_{\phi_1\phi_2\mathchar`-\psi} G_0^{\phi\phi\phi} \nonumber \\
	& = & G_0^{\phi\phi\phi} V_{\psi\mathchar`-\phi_1\phi_2} {\cal G}' \frac{1}{1 - V_{\phi_1\phi_2\mathchar`-\psi} G_0^{\phi\phi\phi} t_3 G_0^{\phi\phi\phi} V_{\psi\mathchar`-\phi_1\phi_2} {\cal G}'} V_{\phi_1\phi_2\mathchar`-\psi} G_0^{\phi\phi\phi},
\end{eqnarray}
where we defined ${\cal G}'$ by
\begin{eqnarray}
	{\cal G}' & = & G_0^{'\psi\phi_3} \left( t_{\psi\phi_3}' + t_{\psi\phi_3}' G_0^{'\psi\phi_3} \left( \Sigma_0 + \Delta + \Delta G^{\psi\phi_3} \Sigma + \Sigma G^{\psi\phi_3} \Delta + \Delta G^{\psi\phi_3} \Delta \right) G_0^{'\psi\phi_3} t_{\psi\phi_3}' + \cdots \right) G_0^{'\psi\phi_3} \nonumber \\
	& = & G_0^{'\psi\phi_3} \left( t_{\psi\phi_3}' + t_{\psi\phi_3}' \left( G^{\psi\phi_3} - G_0^{'\psi\phi_3} - G_0^{'\psi\phi_3} \left( \Sigma_V + \Sigma G^{\psi\phi_3} \Sigma \right) G_0^{'\psi\phi_3} \right) t_{\psi\phi_3}' + \cdots \right) G_0^{'\psi\phi_3}. %\nonumber \\
%	& = & G_0^{'\psi\phi_3} t_{\psi\phi_3}' \frac{1}{1 - \left( G^{\psi\phi_3} - G_0^{'\psi\phi_3} - G_0^{'\psi\phi_3} \left( \Sigma_V + \Sigma G^{\psi\phi_3} \Sigma \right) G_0^{'\psi\phi_3} \right) T^{' \left( 2 \right)}} G_0^{'\psi\phi_3}
\end{eqnarray}
Its diagrammatic expression is given as
\begin{equation}
	\includegraphics[width=18cm]{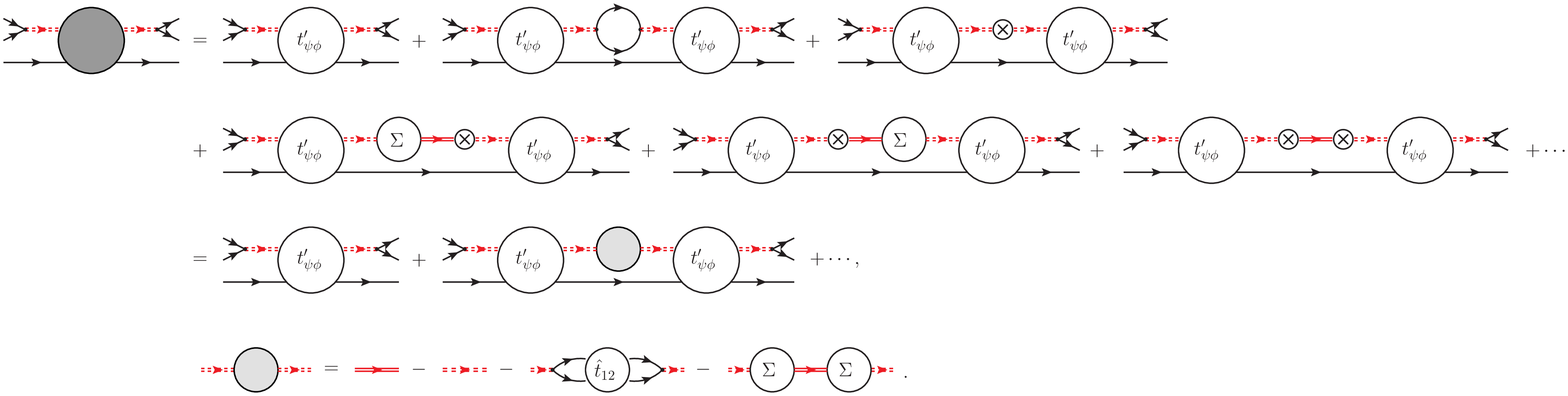}
\end{equation}
With ${\cal G}'$ defined above, $w_3$ is written as
\begin{equation}
	w_3 = G_0^{\phi\phi\phi} V_{\psi\mathchar`-\phi_1\phi_2} {\cal G}' V_{\phi_1\phi_2\mathchar`-\psi} G_0^{\phi\phi\phi} + G_0^{\phi\phi\phi} V_{\psi\mathchar`-\phi_1\phi_2} {\cal G}' V_{\phi_1\phi_2\mathchar`-\psi} G_0^{\left( 3 \right)}\ t_3\ G_0^{\phi\phi\phi} V_{\psi\mathchar`-\phi_1\phi_2} {\cal G}' V_{\phi_1\phi_2\mathchar`-\psi} G_0^{\phi\phi\phi} + \cdots.
\end{equation}
We can see that the counterterms in higher order terms of the effective AGS equations are now put into ${\cal G}'$ while keeping the form of the effective AGS equations the same. \par
\begin{figure}[htbp]
	\includegraphics[width=10cm]{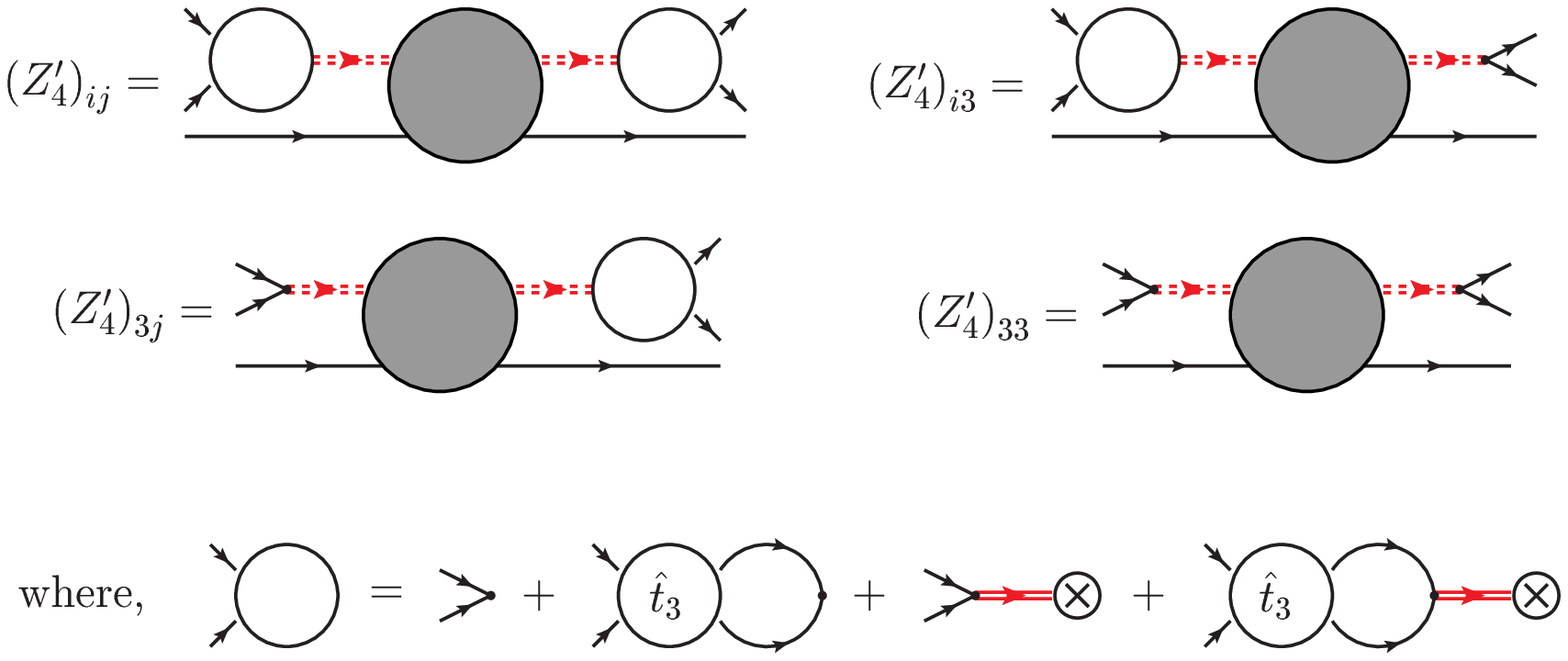}
	\caption{A diagrammatic representation of the modified kernel.}
	\label{fig:modifiedZ}
\end{figure}

We have been focused on the intermediate part of the scattering processes. We next discuss the left- and right-most parts of the scattering processes which have the channel-dependent structure. Explicit expressions for the left- and right-most structures of the last term in $W_3$ are,
\begin{eqnarray}
1 + Z_0 T_3 =  \left(
\begin{array}{ccc}
	1 & 0 & G_0^{\phi\phi\phi} t_3 \\
	0 & 1 & G_0^{\phi\phi\phi} t_3 \\
	0 & 0 & 1 \\
\end{array} 
\right),
\hspace{1cm}
T_3 Z_0 + 1 = \left(
\begin{array}{ccc}
	1 & 0 & 0 \\
	0 & 1 & 0 \\
	G_0^{\phi\phi\phi} t_3 & G_0^{\phi\phi\phi} t_3 & 1 \\
\end{array} 
\right).
\end{eqnarray}
Explicit matrix expressions for $\left( 1 + Z_0 T_3 \right) W_3 \left( T_3 Z_0 + 1 \right)$ is therefore,
\begin{eqnarray}
\left(
\begin{array}{ccc}
	\left( 1 + G_0^{\phi\phi\phi} t_3 \right) w_3 \left( t_3 G_0^{\phi\phi\phi} + 1 \right) & \left( 1 + G_0^{\phi\phi\phi} t_3 \right) w_3 \left( t_3 G_0^{\phi\phi\phi} + 1 \right) & \left( 1 + G_0^{\phi\phi\phi} t_3 \right) w_3 \\
	\left( 1 + G_0^{\phi\phi\phi} t_3 \right) w_3 \left( t_3 G_0^{\phi\phi\phi} + 1 \right) & \left( 1 + G_0^{\phi\phi\phi} t_3 \right) w_3 \left( t_3 G_0^{\phi\phi\phi} + 1 \right) & \left( 1 + G_0^{\phi\phi\phi} t_3 \right) w_3 \\
	w_3 \left( t_3 G_0^{\phi\phi\phi} + 1 \right) & w_3 \left( t_3 G_0^{\phi\phi\phi} + 1 \right) & w_3 \\
\end{array} 
\right).
\end{eqnarray}
Noting that the left- and the right-most factor of $w_3$ is $ G_0^{\phi\phi\phi} V_{\psi\mathchar`-\phi_1\phi_2}$ and $V_{\phi_1\phi_2\mathchar`-\psi} G_0^{\left( 3 \right)}$, we obtain
\begin{eqnarray}
	\left( 1 + G_0^{\phi\phi\phi} t_3 \right) G_0^{\phi\phi\phi} V_{\psi\mathchar`-\phi_1\phi_2} & = & G_0^{\phi\phi\phi} \left( 1 + {\hat t_3} G_0^{\phi\phi\phi} \right) V_{\psi\mathchar`-\phi_1\phi_2} \left( 1 + G^{\psi\phi_3} \Sigma \right) \\
	V_{\phi_1\phi_2\mathchar`-\psi} G_0^{\phi\phi\phi} \left( t_3 G_0^{\phi\phi\phi} + 1 \right) & = & \left( \Sigma G^{\psi\phi_3} + 1 \right) V_{\phi_1\phi_2\mathchar`-\psi} \left( G_0^{\phi\phi\phi} {\hat t_3} + 1 \right) G_0^{\phi\phi\phi}.
\end{eqnarray}
The $\left( 1, 1 \right)$ component, for example, is then written as
\begin{eqnarray}
	& & G_0^{\phi\phi\phi} \left( 1 + {\hat t_3} G_0^{\phi\phi\phi} \right) V_{\psi\mathchar`-\phi_1\phi_2} \left( 1 + G^{\psi\phi_3} \Sigma \right) G_0^{\psi\phi_3} t_{\psi\phi_3}^{\Sigma} G_0^{\psi\phi_3} \left( \Sigma G^{\psi\phi_3} + 1 \right) V_{\phi_1\phi_2\mathchar`-\psi} \left( G_0^{\phi\phi\phi} {\hat t_3} + 1 \right) G_0^{\phi\phi\phi} \nonumber \\
	& = & G_0^{\phi\phi\phi} \left( 1 + {\hat t_3} G_0^{\phi\phi\phi} \right) V_{\psi\mathchar`-\phi_1\phi_2} G^{\psi\phi_3} t_{\psi\phi_3}^{\Sigma} G^{\psi\phi_3} V_{\phi_1\phi_2\mathchar`-\psi} \left( G_0^{\phi\phi\phi} {\hat t_3} + 1 \right) G_0^{\phi\phi\phi}.
\end{eqnarray}
Similarly, $\left( 1, 3 \right)$ component is written as
\begin{equation}
	G_0^{\phi\phi\phi} \left( 1 + {\hat t_3} G_0^{\phi\phi\phi} \right) V_{\psi\mathchar`-\phi_1\phi_2} G^{\psi\phi_3} t_{\psi\phi_3}^{\Sigma} G_0^{\psi\phi_3} V_{\phi_1\phi_2\mathchar`-\psi} G_0^{\phi\phi\phi},
\end{equation}
and $\left( 3, 1 \right)$ component as
\begin{equation}
	G_0^{\phi\phi\phi} V_{\psi\mathchar`-\phi_1\phi_2} G_0 t_{\psi\phi_3}^{\Sigma} G^{\psi\phi_3} V_{\phi_1\phi_2\mathchar`-\psi} \left( G_0^{\phi\phi\phi} {\hat t_3} + 1 \right) G_0^{\phi\phi\phi}.
\end{equation}
Diagrammatic representation of each matrix element is given as
\begin{equation}
	\includegraphics[width=17cm]{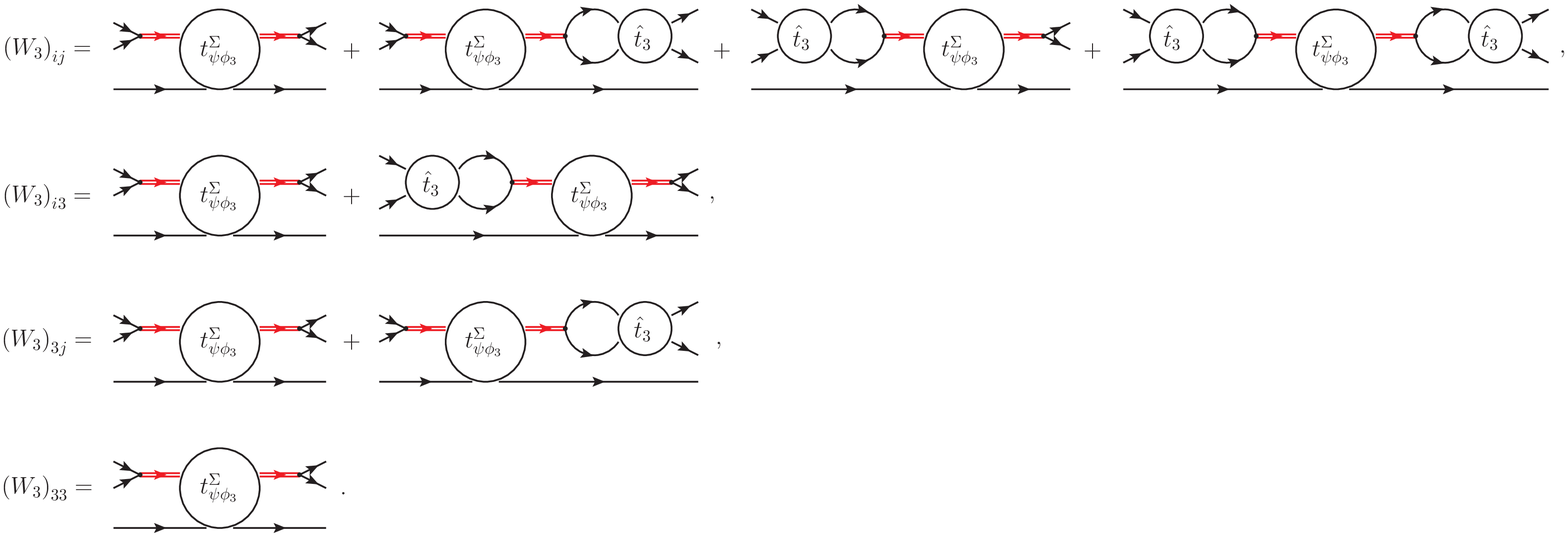}
\end{equation}
We now rewrite them in terms of the renormalized quantities as we did in the previous paragraph. For example, $\left( 1, 3 \right)$ component is written as
\begin{eqnarray}
	& & G_0^{\phi\phi\phi} \left( 1 + {\hat t_3} G_0^{\phi\phi\phi} \right) V_{\psi\mathchar`-\phi_1\phi_2} G^{\psi\phi_3} t_{\psi\phi_3}^{\Sigma} G_0^{\psi\phi_3} V_{\phi_1\phi_2\mathchar`-\psi} G_0^{\phi\phi\phi} \nonumber \\
	& = & G_0^{\phi\phi\phi} \left( 1 + {\hat t_3} G_0^{\phi\phi\phi} \right) V_{\psi\mathchar`-\phi_1\phi_2} \left( 1 + G^{\psi\phi_3} \left( \Sigma + \Delta \right) \right) G_0^{'\psi\phi_3} t_{\psi\phi_3}^{\Sigma} G_0^{\psi\phi_3} V_{\phi_1\phi_2\mathchar`-\psi}.
\end{eqnarray}
We include the counterterm by modifying the left-most part of the kernel as
\begin{equation}
	G_0^{\phi\phi\phi} \left( 1 + {\hat t_3} G_0^{\phi\phi\phi} \right) V_{\psi\mathchar`-\phi_1\phi_2} \left( 1 + G^{\psi\phi_3} \Delta \right) {\cal G}' V_{\phi_1\phi_2\mathchar`-\psi} G_0^{\phi\phi\phi}.
\end{equation}
For $\left( 3, 1 \right)$ component, we modify the kernel as
\begin{equation}
	G_0^{\phi\phi\phi} V_{\psi\mathchar`-\phi_1\phi_2} {\cal G}' \left( \Delta G^{\psi\phi_3} + 1 \right) V_{\phi_1\phi_2\mathchar`-\psi} \left( G_0^{\phi\phi\phi} {\hat t_3} + 1 \right) G_0^{\phi\phi\phi},
\end{equation}
and for $\left( 1, 1 \right)$ component as
\begin{equation}
G_0^{\phi\phi\phi} \left( 1 + {\hat t_3} G_0^{\phi\phi\phi} \right) V_{\psi\mathchar`-\phi_1\phi_2} \left( 1 + G^{\psi\phi_3} \Delta \right) {\cal G}' \left( \Delta G^{\psi\phi_3} + 1 \right) V_{\phi_1\phi_2\mathchar`-\psi} \left( G_0^{\phi\phi\phi} {\hat t_3} + 1 \right) G_0^{\phi\phi\phi}.
\end{equation}
Each component of the modified driving term is as follows
\begin{eqnarray}
\begin{array}{rcl}
	\left( Z_4' \right)_{ij}  & = & G_0^{\phi\phi\phi} \left( 1 + {\hat t_3} G_0^{\phi\phi\phi} \right) V_{\psi\mathchar`-\phi_1\phi_2} \left( 1 + G^{\psi\phi_3} \Delta \right) {\cal G}' \left( \Delta G^{\psi\phi_3} + 1 \right) V_{\phi_1\phi_2\mathchar`-\psi} \left( G_0^{\phi\phi\phi} {\hat t_3} + 1 \right) G_0^{\phi\phi\phi} \hspace{0.3cm} (i, j = 1, 2) \\
	\left( Z_4' \right)_{i3} & = & G_0^{\phi\phi\phi} \left( 1 + {\hat t_3} G_0^{\phi\phi\phi} \right) V_{\psi\mathchar`-\phi_1\phi_2} \left( 1 + G^{\psi\phi_3} \Delta \right) {\cal G}' V_{\phi_1\phi_2\mathchar`-\psi} G_0^{\phi\phi\phi} \hspace{0.3cm} (i = 1, 2) \\
	\left( Z_4' \right)_{3j} & = & G_0^{\phi\phi\phi} V_{\psi\mathchar`-\phi_1\phi_2} {\cal G}' \left( \Delta G^{\psi\phi_3} + 1 \right) V_{\phi_1\phi_2\mathchar`-\psi} \left( G_0^{\phi\phi\phi} {\hat t_3} + 1 \right) G_0^{\phi\phi\phi} \hspace{0.3cm} (j = 1, 2) \\
	\left( Z_4' \right)_{33} & = & G_0^{\phi\phi\phi} V_{\psi\mathchar`-\phi_1\phi_2} {\cal G}' V_{\phi_1\phi_2\mathchar`-\psi} G_0^{\phi\phi\phi}.
\end{array}
\end{eqnarray}
Diagrammatic representation of each component is given as in Fig.\ref{fig:modifiedZ}.

\section{Jacobi Momenta, Explicit Expressions for the Matrix Elements in the Effective AGS Equations}
\label{sec:appendix2}

\subsection{Jacobi Momenta}
In a three-body system, it is convenient to introduce the Jacobi momenta defined by
\begin{eqnarray}
	{\bf P} & = & {\bf k}_1 + {\bf k}_2 + {\bf k}_3, \\
	{\bf q}_i & = & \frac{ \left( m_j + m_k \right) {\bf k}_i - m_i \left( {\bf k}_j + {\bf k}_k \right) }{m_i + m_j + m_k}, \\
	{\bf p}_i & = & \frac{ m_k {\bf k}_j - m_j {\bf k}_k }{m_j + m_k},
\end{eqnarray}
where $i, j, k$ are cyclic permutation of $1, 2, 3$ and ${\bf k}_i$ is a momentum of $\phi_i$ in Cartesian coordinate. In the following, we consider the problem in the three-body center-of-mass frame, that is, we set
\begin{equation}
	{\bf P} = {\bf 0}.
\end{equation}
Other two Jacobi momenta then become
\begin{eqnarray}
	{\bf q}_i & = & {\bf k}_i, \\
	{\bf p}_i & = & \frac{ m_k {\bf k}_j - m_j {\bf k}_k }{m_j + m_k}.
\end{eqnarray}
We also introduce reduced masses as follows
\begin{eqnarray}
	M_i^{-1} & = & \left( m_j + m_k \right)^{-1} + m_i^{-1} \\
	\mu_{\phi_j\phi_k}^{-1} & = & m_j^{-1} + m_k^{-1}.
\end{eqnarray}
With these notation we introduced, the kinetic energy in the three-body channel in the center-of-mass frame is written as
\begin{eqnarray}
	& & \sum_{i=1}^3 \left( m_i + \frac{k_i^2}{2 m_i} \right) \nonumber \\
	& = & \sum_{i=1}^3 m_i + \frac{P^2}{2 \sum_{i=1}^3 m_i} + \frac{q_i^2}{2 M_i} + \frac{p_i^2}{2 \mu_{\phi_j\phi_k}} \nonumber \\
	& = & \sum_{i=1}^3 m_i + \frac{q_i^2}{2 M_i} + \frac{p_i^2}{2 \mu_{\phi_j\phi_k}}.
\end{eqnarray}
\par
We denote the center-of-mass and the relative momentum in the two-body channel as
\begin{eqnarray}
	{\cal P} & = & {\bf K}_3 + {\bf k}_3, \\
	{\bf p} & = & \frac{m_3 {\bf K}_3 - M {\bf k}_3}{M + m_3}.
\end{eqnarray}
We denote the reduced mass of $\psi_3\phi_3$ as $\mu_{\psi\phi_3}$
\begin{equation}
	\left( \mu_{\psi\phi_3} \right)^{-1} = M^{-1} + m_3^{-1},
\end{equation}
where we placed superscript on the mass of $\psi_3$ since its mass shifts from the bare one by the coupling to $\phi_1\phi_2$ two-body state. The kinetic energy in the two-body channel in the center-of-mass frame is therefore written as
\begin{eqnarray}
	& & M + \frac{K_3^2}{2 M} + m_3 + \frac{k_3^2}{2 m_3} \nonumber \\
	& = & M + m_3 + \frac{{\cal P}^2}{2 \left( M + m_3 \right)} + \frac{p^2}{2 \mu_{\psi\phi_3}} \nonumber \\
	& = & M + m_3 + \frac{p^2}{2 \mu_{\psi\phi_3}}.
\end{eqnarray}
In section (\ref{numericalresults}), we perform numerical calculation in a case where the two-body and three-body thresholds are degenerate
\begin{equation}
	M' = m_1 + m_2,
\end{equation}
where we introduced the physical mass of $\psi_3$, $M'$. We briefly discuss how each kinetic quantities are related in that case. If two thresholds are degenerate, we have
\begin{equation}
	{\bf K}_3 = {\bf k}_1 + {\bf k}_2,
\end{equation}
which leads to
\begin{equation}
	{\bf P} = {\bf k}_1 + {\bf k}_2 + {\bf k}_3 = {\bf K}_3 + {\bf k}_3 = {\cal P}.
\end{equation}
That is, the three-body center-of-mass frame is also the two-body center-of-mass frame. The relative momentum between the pair $\phi_1\phi_2$ and $\phi_3$ is then
\begin{eqnarray}
	{\bf q}_3 & = & \frac{\left( m_1 + m_2 \right) {\bf k}_3 - m_3 \left( {\bf k}_1 + {\bf k}_2 \right)}{ m_1 + m_2 + m_3 } \nonumber \\
	& = & \frac{M' {\bf k}_3 - m_3 {\bf K}_3}{M' + m_3} = - {\bf p}.
\end{eqnarray}
In a word, the relative momentum between $\phi_3$ and $\phi_1\phi_2$ is equal (up to sign) to the relative momentum between $\psi_3$ and $\phi_3$.

\subsection{Elementary, Effective Interactions, $T$-Matrices and Kernels}

 In this subsection, we present explicit expressions for elementary and the effective interactions and its matrix elements. Matrix elements of $Z_0 \left( E \right)$ and (unmodified) $Z_4 \left( E \right)$ are also presented. \par
Elementary interactions between $\phi_j\phi_k$ is of the form
\begin{equation}
	V_{\phi_j\phi_k} = \int^{\infty}_0 q_i^2 dq_i | q_i g \rangle \lambda_{\phi_j\phi_k} \langle q_i g |,
\end{equation}
and those couple $\psi$ and $\phi_1\phi_2$ are,
\begin{equation}
	V_{\psi\mathchar`-\phi_1\phi_2} = \int^{\infty}_0 q_3^2 dq_3 \Gamma | q_3 \rangle \langle q_3 g |,
\end{equation}
and an interaction between $\psi\phi_3$ is,
\begin{equation}
	V_{\psi\phi_3} = | g \rangle \lambda_{\psi\phi_3} \langle g |,
\end{equation}
where $| g \rangle$ are Yamaguchi-type form factor.
The effective interactions are therefore
\begin{eqnarray}
	U_i \left( E \right) & = & V_{\phi_j\phi_k} = \int^{\infty}_0 q_i^2 dq_i | q_i g \rangle \lambda_{\phi_j\phi_k} \langle q_i g | \hspace{1cm} (i = 1, 2), \\
	U_3 \left( E \right) & = & V_{\phi_1\phi_2} + V_{\psi\mathchar`-\phi_1\phi_2} G_0^{\psi\phi_3} \left( E \right) V_{\phi_1\phi_2\mathchar`-\psi} = \int^{\infty}_0 q_3^2 dq_3 | q_3 g \rangle \left( \lambda_{\phi_1\phi_2} + \Gamma G_0^{\psi\phi_3} \left( E q_3 \right) \Gamma \right) \langle q_3 g |, \\
	U_4 \left( E \right) & = & V_{\psi\mathchar`-\phi_1\phi_2} G_0^{\psi\phi_3} \left( E \right) t_{\psi\phi_3} \left( E \right) G_0^{\psi\phi_3} \left( E \right) V_{\phi_1\phi_2\mathchar`-\psi} \nonumber \\
	& = & \int^{\infty}_0 q_3^2 dq_3 q_3'^2 dq_3' \Gamma | q_3 g \rangle G_0^{\psi\phi_3} \left( E q_3 \right) \langle q_3 | t_{\psi\phi_3} \left( E \right) | q_3' \rangle G_0^{\psi\phi_3} \left( E q_3' \right) \langle q_3' g | \Gamma.
\end{eqnarray}
As we saw in the previous appendix, when it comes to consider the modified kernel in the presence of the elementary interactions between $\phi_1\phi_2$, it is convenient to express the two-body $T$-matrix of $\phi_1\phi_2$ as follows
\begin{equation}
	t_3 \left( E \right) = {\hat t_3} \left( E \right) + \left( 1 + {\hat t_3} \left( E \right) G_0^{\phi\phi\phi} \left( E \right) \right) V_{\psi\mathchar`-\phi_1\phi_2} G^{\psi\phi_3} \left( E \right) V_{\phi_1\phi_2\mathchar`-\psi} \left( G_0^{\phi\phi\phi} \left( E \right) {\hat t_3} \left( E \right) + 1 \right),
\end{equation}
which becomes
\begin{equation}
	t_3 \left( E \right) = \int q_3^2 dq_3 | q_3 g \rangle \left( {\hat \tau_3} \left( E q_3 \right) + \left( 1 + {\hat \tau_3} \left( E q_3 \right) {\cal L}^{\left( 3 \right)} \left( E q_3 \right) \right) \Gamma G^{\psi\phi_3} \left( E q_3 \right) \Gamma \left( {\cal L}^{\left( 3 \right)} \left( E q_3 \right) {\hat \tau_3} \left( E q_3 \right) + 1 \right) \right) \langle q_3 g |.
\end{equation}
We defined a one-loop integral in three-body channel
\begin{equation}
	{\cal L}^{\left( 3 \right)} \left( E q_3 \right) = \int p_3^2 dp_3\ g^2 \left( p_3 \right) G_0^{\phi\phi\phi} \left( E q_3 p_3 \right),
\end{equation}
where $| g \rangle$ is Yamaguchi-type form factor
\begin{equation}
	\langle p | g \rangle = g \left( p \right) = \frac{\Lambda^2}{p^2 + \Lambda^2},
\end{equation}
and we introduced the two-body Green function
\begin{equation}
	G_0^{\psi\phi_3} \left( E q_3 \right) = \left( E - M' - m_3 - \frac{ q_3^2 }{2 \mu_{\psi\phi_3}} \right)^{-1}.
\end{equation}
Since $V_{\psi\phi_3}$ is separable, the $T$-matrix is expressed as follows
\begin{equation}
	t_{\psi\phi_3} \left( E \right) = | g \rangle \tau_{\psi\phi_3} \langle g |, \hspace{1cm}
	\left( \tau_{\psi\phi_3} \left( E \right) \right)^{-1} = \left( \lambda_{\psi\phi_3} \right)^{-1} - \frac{\pi}{2} \frac{\mu_{\psi\phi_3} \Lambda^3}{\left( k + i \Lambda \right)^2},
\end{equation}
where $k$ is the relative momentum between $\psi_3$ and $\phi_3$ defined by below
\begin{equation}
	E = M' + m_3 + \frac{k^2}{2 \mu_{\psi\phi_3}}.
\end{equation}
It is convenient to define dimensionless coupling constant which we denote as $f_{\psi\phi_3}$
\begin{equation}
	f_{\psi\phi_3} = \lambda_{\psi\phi_3} \frac{\pi}{2} \mu_{\psi\phi_3} \Lambda.
\end{equation}
$t_{\psi\phi_3} \left( E \right)$ has a bound state pole for $f_{\psi\phi_3} \leq -1$ and a virtual state pole for $f_{\psi\phi_3} \geq -1$. The effective three-body interaction $U_4 \left( E \right)$ is then written as
\begin{equation}
	U_4 \left( E \right) = \int^{\infty}_0 q_3^2 dq_3 q_3'^2 dq_3' \Gamma | q_3 g \rangle G_0^{\psi\phi_3} \left( E q_3 \right) g \left( q_3 \right) \tau_{\psi\phi_3} \left( E \right) g \left( q_3' \right) G_0^{\psi\phi_3} \left( E q_3' \right) \langle q_3' g | \Gamma.
\end{equation}
Since we assume separable interactions for two-body interactions among $\phi_1\phi_2\phi_3$, the two-body $T$-matrices are expressed as follows
\begin{equation}
	t_i \left( E \right) = \int^{\infty}_0 q_i^2 dq_i | q_i g \rangle \tau_i \left( E q_i \right) \langle q_i g |,
\end{equation}
where
\begin{equation}
	\left( \tau_i \left( E q_i \right) \right)^{-1} = U_i^{-1} \left( E q_i \right) - \int^{\infty}_0 p_i^2 dp_i\ g^2 \left( p_i \right) G_0^{\phi\phi\phi} \left( E q_i p_i \right) = U_i^{-1} \left( E q_i \right) - \frac{\pi}{2} \frac{\mu_{\phi_j\phi_k} \Lambda^3}{ \left( k_i \left( E q_i \right) + i \Lambda \right)^2 },
\end{equation}
and
\begin{equation}
	E = m_i + m_j + m_k + \frac{k_i^2 \left( E q_i \right)}{2 \mu_{\phi_j\phi_k}} + \frac{q_i^2}{2 M_i}.
\end{equation}
$k_i$ is clearly the relative momentum between $\phi_j\phi_k$. We also present dimensionless coupling constants in the three-body channels as follows
\begin{equation}
	f_{\phi_j\phi_k} = \lambda_{\phi_j\phi_k} \frac{\pi}{2} \mu_{\phi_j\phi_k} \Lambda.
\end{equation}
The three-body $T$-matrix is written as follows
\begin{eqnarray}
	t_4 \left( E \right) & = & U_4 \left( E \right) + U_4 \left( E \right) G_0^{\phi\phi\phi} \left( E \right) U_4 \left( E \right) + \cdots \nonumber \\
	& = & \int q_3^2 dq_3 q_3'^2 dq_3'\ \Gamma | q_3 g \rangle \nonumber \\
	& \times & G_0^{\psi\phi_3} \left( E q_3 \right) g \left( q_3 \right) \tau_{\psi\phi_3} \frac{1}{1 - \langle g | G_0^{\psi\phi_3} \left( E \right) \Sigma_0 \left( E \right) G_0^{\psi\phi_3} \left( E \right) | g \rangle \tau_{\psi\phi_3} \left( E \right)} g \left( q_3' \right) G_0^{\psi\phi_3} \left( E q_3' \right) \nonumber \\
	& \times & \langle q_3' g | \Gamma.
\end{eqnarray}
The three-body $T$-matrix is written as follows
\begin{eqnarray}
	t_4 \left( E \right) & = & U_4 \left( E \right) + U_4 \left( E \right) G_0^{\phi\phi\phi} \left( E \right) U_4 \left( E \right) + \cdots \nonumber \\
	& = & \int q_3^2 dq_3 q_3'^2 dq_3'\ \Gamma | q_3 g \rangle G_0^{\psi\phi_3} \left( E q_3 \right) g \left( q_3 \right) \tau_{\psi\phi_3} \frac{1}{1 - {\cal V} \left( E \right) \tau_{\psi\phi_3} \left( E \right)} g \left( q_3' \right) G_0^{\psi\phi_3} \left( E q_3' \right) \langle q_3' g | \Gamma, \nonumber
\end{eqnarray}
where we defined
\begin{equation}
	{\cal V} \left( E \right) = \langle g | G_0^{\psi\phi_3} \left( E \right) \Sigma_0 \left( E \right) G_0^{\psi\phi_3} \left( E \right) | g \rangle.
\end{equation}

\subsection{The Unmodified and the Modified Kernel}
The unmodified kernels of the AGS equations are therefore
\begin{eqnarray}
	Z_0 \left( E \right) & = & G_0^{\phi\phi\phi} \left( E \right) {\bar \delta} \\
	Z_4 \left( E \right) & = & G_0^{\phi\phi\phi} \left( E \right) t_4 \left( E \right) {\bf 1} G_0^{\phi\phi\phi} \left( E \right) \\
	K \left( E \right) & = & \left( Z_0 \left( E \right) + Z_4 \left( E \right) \right) T \left( E \right).
\end{eqnarray}
The modified $Z_4 \left( E \right)$ are obtained as follows. First, we replace ${\cal V}\left( E \right)$ with
\begin{equation}
	\langle g | G_0^{\psi\phi_3} \left( E \right) \left( \Sigma \left( E \right) + \Sigma \left( E \right) G^{\psi\phi_3} \left( E \right) \Delta + \Delta G^{\psi\phi_3} \left( E \right) \Sigma \left( E \right) + \Delta G^{\psi\phi_3} \left( E \right) \Delta \right) G_0^{\psi\phi_3} \left( E \right) | g \rangle,
\end{equation}
which can also be written as follows
\begin{equation}
	\langle g | G^{\psi\phi_3} \left( E \right) - G_0^{\psi\phi_3} \left( E \right) - G_0^{\psi\phi_3} \left( E \right) \Sigma \left( E \right) G^{\psi\phi_3} \left( E \right) \Sigma \left( E \right) G_0^{\psi\phi_3} \left( E \right) | g \rangle.
\end{equation}
For notational simplicity, we denote the above quantity as ${\cal V} \left( E \right)$. Second, we also replace the left-most structure with the following
\begin{equation}
	V_{\psi\mathchar`-\phi_1\phi_2} \Rightarrow V_{\psi\mathchar`-\phi_1\phi_2} + {\bar \delta} I_3 \left( 1 + {\hat t_3} \left( E \right) G_0^{\phi\phi\phi} \left( E \right) \right) V_{\psi\mathchar`-\phi_1\phi_2} G^{\psi\phi_3} \left( E \right) \Delta	= V^{\left( 32 \right)'},
\end{equation}
and the right-most structure with the corresponding one
\begin{equation}
	V_{\phi_1\phi_2\mathchar`-\psi} \Rightarrow \Delta G^{\psi\phi_3} \left( E \right) V^{\left(23 \right)} \left( G_0^{\phi\phi\phi} \left( E \right) {\hat t_3} \left( E \right) + 1\right) I_3 {\bar \delta} + V_{\psi\mathchar`-\phi_1\phi_2} = V^{\left( 23 \right)'}.
\end{equation}
An explicit expression for the modified driving term $Z_4' \left( E \right)$ is therefore obtained as follows
\begin{equation}
	Z_4' \left( E \right) = \int q_3^2 dq_3 q_3'^2 dq_3'\ G_0^{\phi\phi\phi} \left( E \right) V^{\left( 32 \right)'} G_0^{\psi\phi_3} \left( E \right) | g \rangle \tau_{\psi\phi_3} \left( E \right) \frac{1}{1 - {\cal V} \left( E \right)} \langle g | {\bf 1} G_0^{\psi\phi_3} \left( E \right)  V^{\left( 23 \right)'} G_0^{\phi\phi\phi} \left( E \right).
\end{equation}

\subsubsection{Matrix Elements of the Driving Terms}
In this subsection, we present explicit expression of the matrix elements of the kernel. $Z_0 \left( E q_i q_j \right)$ is represented as follows
\begin{equation}
	Z_0 \left( E q_i q_j \right) = \langle q_j g | {\bar \delta}_{ij} G_0^{\phi\phi\phi} \left( E \right) | q_j g \rangle = \frac{{\bar \delta}_{ij}}{2} \int^1_{-1} dx \frac{ g \left( p_i \right) g \left( p_j \right) }{ E - \frac{q_i^2}{2 m_i} - \frac{q_j^2}{2 m_j} - \frac{ \left( {\bf q}_i + {\bf q}_j \right)^2 }{2 m_k} },
\end{equation}
where $x$ and ${\bar \delta}_{ij}$ are defined by below
\begin{equation}
	x = {\bf {\hat q}_i} \cdot {\bf {\hat q}_j}, \hspace{0.5cm} {\bar \delta}_{ij} = 1 - \delta_{ij}.
\end{equation}
$p_i$ and $p_j$ are absolute value of relative momenta between $\phi_j\phi_k$ and $\phi_k\phi_i$ respectively whose explicit expressions are
\begin{equation}
	p_i = \left| - \frac{ m_j }{ m_j + m_k } {\bf q_i} - {\bf q_j} \right|, \hspace{1cm} p_j = \left| {\bf q_i} + \frac{ m_k }{ m_k + m_i } {\bf q_j} \right|.
\end{equation}
Matrix elements of the unmodified $Z_4 \left( E q_3 q_3' \right)$ is given as
\begin{eqnarray}
	Z_4 \left( E q_3 q_3' \right) & = & \langle q_3 g | G_0^{\phi\phi\phi} \left( E \right) t_4 \left( E \right) G_0^{\phi\phi\phi} \left( E \right) | q_3' g \rangle \nonumber \\
	& = & \int p_3^2 dp_3\ g\left( p_3 \right) G_0^{\phi\phi\phi} \left( E q_3 p_3 \right) \Gamma g \left( p_3 \right)\ G_0^{\psi\phi_3} \left( E q_3 \right) g \left( q_3 \right) \tau_{\psi\phi_3} \left( E \right) \frac{1}{1 - {\cal V} \left( E \right) \tau_{\psi\phi_3} \left( E \right)} g \left( q_3' \right) G_0^{\psi\phi_3} \left( E q_3' \right) \nonumber \\
	& \times & \int p_3'^2 dp_3'\ g \left( p_3' \right) \Gamma G_0^{\phi\phi\phi} \left( E q_3 p_3 \right) g \left( p_3' \right).
\end{eqnarray}
Other matrix elements of the unmodified $Z_4 \left( E \right)$ are obtained in a straightforward way. \par
As we saw in the previous section, we need to perform the reorganization of the AGS equations and one of the kernel $Z_4 \left( E \right)$ need to be replaced by the modified one. ${\cal V} \left( E \right)$ is modified as
\begin{equation}
	{\cal V} \left( E \right) = \int q_3^2 dq_3 \left( G^{\psi\phi_3} \left( E q_3 \right) - G_0^{\psi\phi_3} \left( E q_3 \right) -  G_0^{\psi\phi_3} \left( E q_3 \right) \left( \Sigma_V \left( E \right) + \Sigma \left( E \right) G_{\Sigma} \left( E \right) \Sigma \left( E \right) \right) G_0^{\psi\phi_3} \left( E q_3 \right) \right).
\end{equation}
For matrix elements $Z_4 \left( E q_i q_j \right)\ i = 1, 2$, the left-most structure of $Z_4 \left( E \right)$ are modified as
\begin{equation}
	\int^{\infty}_0 p_3^2 dp_3\ g \left( p_3 \right) G_0^{\phi\phi\phi} \left( E q_3 p_3 \right) \left( \Gamma g \left( p_3 \right) + \left( 1 + {\hat \tau_3} \left( E \right) G_0^{\phi\phi\phi} \left( E q_3 p_3 \right) \right) \Gamma g \left( p_3 \right) G^{\psi\phi_3} \left( E q_3 \right) \Delta \right).
\end{equation}
Similarly, for matrix elements $Z_4 \left( E q_i q_j \right)\ j = 1, 2$ the right-most parts are modified as
\begin{equation}
	\int^{\infty}_0 p_3^2 dp_3 \left( g \left( p_3 \right) \Gamma + \Delta G^{\psi\phi_3} \left( E q_3 \right) g \left( p_3 \right) \Gamma \left( G_0^{\phi\phi\phi} \left( E q_3 p_3 \right) {\hat \tau_3} \left( E \right) + 1 \right) \right) G_0^{\phi\phi\phi} \left( E q_3 p_3 \right) g \left( p_3' \right).
\end{equation}

\bibliography{ref.bib}

\end{document}